\documentclass{article}
\usepackage{amssymb,amsfonts,amsmath}
\usepackage{cite,enumerate,float}
\usepackage{color}
\usepackage{tikz}
\usetikzlibrary{calc}
\usetikzlibrary{arrows,snakes,backgrounds}
\DeclareGraphicsExtensions{.pdf,.png,.jpg}
\usepackage{multirow}
\usepackage{pgfplots}
\usetikzlibrary{arrows.meta,positioning}
\usetikzlibrary{arrows.meta,positioning,calc}
\pgfplotsset{compat=1.9}
\usepackage{amsmath, amsfonts, amssymb, amsthm, mathtools}
\usepackage{rotating}
\usepackage{ytableau}
\usepackage{tcolorbox}
 \ytableausetup{mathmode, boxsize=0.4 em}
\usepackage{braket}
\usepackage{float}
\usepackage{cancel}
\usepackage{amsmath}
\usepackage{tikz}
\usetikzlibrary{calc,matrix}
\usepackage{wrapfig}
\usepackage{graphicx}
\graphicspath{{pictures/}}
\usepackage{hyperref}
\usepackage{bm}
\DeclareGraphicsExtensions{.pdf,.png,.jpg}
\usepackage{array}
\usepackage[T2A]{fontenc}
\usepackage[utf8]{inputenc}
\def\be{\begin{eqnarray}}
\def\ee{\end{eqnarray}}

\definecolor{red}{rgb}{1,0,0}
\definecolor{orange}{rgb}{1,0.5,0}
\definecolor{violet}{rgb}{0.7,0,1}

\hoffset= - 1.0in

\begin{document}

\begin{center}
MIPT/TH-22/25 \hfill {\bf In memory of Md Abdur Rashid,}\\
ITEP/TH-30/25 \hfill {\bf with gratitude and sorrow}\\
IITP/TH-27/25 \hfill \phantom{.}\\
\end{center}

\bigskip

\centerline{\Large\textbf{Conformal blocks of Wess-Zumino-Witten model }}
\vspace{4pt}
\centerline{\Large\textbf{from its free-field representation}}

\bigskip
\renewcommand{\thefootnote}{\fnsymbol{footnote}}
\centerline{\textbf{Alexei Morozov$^{1,2,3,4}$\footnote[2]{morozov@itep.ru} \ and\ \   Hasib Sifat$^{1,5}$\footnote[3]{hasib.sifat@math.msu.ru}}}

\bigskip

\centerline{\it $^1$Moscow Institute of Physics and Technology, 141700, Dolgoprudny, Russia}

\centerline{\it $^2$NRC “Kurchatov Institute”, 123182, Moscow, Russia}

\centerline{\it $^3$Institute for Theoretical and Experimental Physics, 117218, Moscow, Russia}

\centerline{\it $^4$Institute for Information Transmission Problems, 127051, Moscow, Russia}

\centerline{\it $^5$Institute for Theoretical and Mathematical Physics,}
\centerline{ \it Lomonosov Moscow State University, 119991 Moscow, Russia}

\bigskip

\bigskip

\centerline{ABSTRACT}

\bigskip

{\footnotesize
A powerful approach to the celebrated Wess-Zumino-Witten (WZW) model is provided by its free-field realization. However, explicit calculations of conformal blocks are not described in the literature in full detail. We begin this study with the simplest cases of the $\hat{sl}(2)_k$ and $\hat{sl}(3)_k$ WZW models, with special emphasis on their global $sl(2)$ and $sl(3)$ symmetries of the resulting correlators, which are not explicit in this formalism. Also non-trivial is the verification of the Knizhnik-Zamolodchikov equations in the $\hat{sl}(3)_k$ case, where the answers take the form of double integrals over screening charge positions and do not look like ordinary hypergeometric functions. }
\tableofcontents

\section{Introduction}

A central problem in the study of any 2d conformal field theory is the computation of correlation functions. In the context of the Wess-Zumino-Witten (WZW) model \cite{WittenWZW}, these correlation functions are  constrained by the infinite-dimensional current algebra. Specifically, the $n$-point correlation functions of primary fields satisfy a system of linear partial differential equations known as the Knizhnik-Zamolodchikov (KZ) equations \cite{KZ}. The space of solutions of KZ equation is finite–dimensional, and a convenient basis of that space are called conformal blocks. Then any correlator of primary fields can be expressed as a linear combination of such conformal blocks. The true correlator is a real-valued bilinear combination of holomorphic and anti-holomorphic blocks. Moreover, KZ equation and its solutions has a very significant importance from the pure mathematical point of view, independently of their original derivation from the conformal Ward identities of the WZW model. They are closely related to study of representations of braid groups, quantum groups and appeared in the context of quantum Langlands program \cite{TK87, EtingofGeer, AFO}. However, solving the KZ equation is not the only way to compute such holomorphic conformal blocks. An alternative and very powerful approach based on free-field representations \cite{DF,DF2,GM} was first developed for minimal models \cite{FF,BPZ,Francesco}
and was later extended and adapted to WZW theories \cite{Wak,Morsl3,GMMOS,FeiginFrenkel,GMM1,GMM2}.
To set the stage of this study, we start with a short technical review of the WZW model and its holomorphic conformal blocks.\\
\newline
WZW model is a two-dimensional conformal field theory with Lie algebra symmetry which is completely solvable. Fields in this theory describes maps from a Riemann surface into a Lie group $ g(z,\bar{z}) : S^2 \to G$. Where $S^2$ is the Riemann sphere, which can be thought as a boundary of a three dimensional ball $B$ ($\partial B = S^2$). Then, the action of this theory is given by \cite{WittenWZW}
\begin{equation}
    S_{WZW}[g] = \frac{k}{4\pi}\left(
    \int_{S^2} d^2z \ \text{tr}(g^{-1}\partial_\mu g g^{-1} \partial_{\mu}g) - \frac{i}{3 } \int_B d^3 y \ \epsilon_{\alpha \beta \gamma} \ \text{tr} (g^{-1}\partial_\mu^\alpha g g^{-1} \partial_{\mu}^\beta g g^{-1} \partial_{\mu}^\gamma g)
    \right)
\label{WZWaction}
\end{equation}
where $k$ is a positive integer. The equation of motion implies the conservation of the chiral current
\begin{equation}
    J(z) = - kJ_z(z) = -k\partial_z gg^{-1}
\end{equation}
The operator product expansion (OPE) in this theory is
just the Kac-Moody algebra
\begin{equation}
    J^a (z) J^b(\omega) \sim \frac{k\delta^{ab}}{(z-\omega)^2} + \frac{i f^{ab}_c J^c(\omega)}{z-\omega}
\end{equation}
Holomorphic $J^a(z)$ is expanded as
\begin{equation}
    J^a(z) = \sum_{n\in \mathbb{Z}} J^a_n z^{-n-1}
\end{equation}
then the modes commute in the conventional way:
\begin{equation}
    [J^a_n, J^b_m] = km\delta^{ab}\delta_{m+n,0}+ i f^{ab}_c J^c_{m+n}
\end{equation}
The term  $km\delta^{ab}\delta_{m+n,0}$ describes the central extension of the loop algebra $\hat G$.

\bigskip
Thus, the WZW model provides the simplest physical realization of Kac-Moody algebras
and their Krichever-Novikov versions \cite{KN,MorKN} for expansions in more sophisticated basises -- which are natural, for example, on non-trivial Riemann surfaces. This is because the non-local term in the action (\ref{WZWaction}) corrects the sigma-model equation of motion from Laplace to holomorphic, what is essential to making the theory conformal, exactly solvable at the quantum level and possessing a free-field realization, typical for generic conformal theories, at least in $2d$.\\

The main property of Wess-Zumino model -- in addition to holomorphicity -- is the Sugawara embedding \cite{Sug} of Virasoro algebra into the square of the Kac-Moody one. Namely the WZW stress tensor
\be
T(z) = \frac{1}{2(k+h^{\wedge})} \sum_a: J^a J^a : (z)
\ee
is holomoprphic and satisfies
\begin{equation}
    T(z)T(\omega) = \frac{c/2}{(z-\omega)^4} + \frac{2 T(\omega)}{(z-\omega)^2} + \frac{\partial T(\omega)}{(z-\omega)}
\end{equation}
where the Virasoro central charge $c$ is expressed through the Kac-Moody level $k$:

\begin{equation}
    c = \frac{k \ dim_{ \mathfrak{g}}}{k + h^{\wedge}}
\end{equation}
where $h^{\wedge}$ is the dual Coxeter number.\\
\bigskip

The free field representations express all the relevant operators in the theory through free fields. In particular case of $2d$ bosonization these are scalars $\phi(z,\bar z)$ with the action
\be
S = \frac{1}{2\pi} \int d^2z \left(\partial \phi \bar\partial \phi + Q R\phi\right)
\label{freef}
\ee
where $R$ is the two dimensional curvature, and the stress tensor
\be
T_{\phi} = - \frac{1}{2} (\partial\phi)^2  - Q\partial^2\phi
\ee
which is holomorphic and has Virasoro central charge $1 - 24 Q^2$.
Vertex operators -- the eigenfunctions of the quadratic stress tensor -- have the form of exponentials:
\begin{equation}
    V_{\alpha(z)} = e^{\alpha \phi(z)}
\end{equation}
The concrete values of the $U(1)$-charges $\alpha$ depend on specific operators in the specific model. The term with $R$ in (\ref{freef}) leads to additional insertions of ``screening charges'' with  $V$ of dimension one, which should be integrated over non-contractible contours along the punctured Riemann surface -- what provides the answers for holomorphic correlators (conformal blocks) in a peculiar form of multiple hypergeometric integrals.\\

Such representation was first studied in detail by V.Dotsenko and V.Fateev \cite{DF} in the case of Virasoro minimal models \cite{BPZ}.
They calculated the 4-point correlator in Virasoro minimal model,
which is now known as the DF partition function
\begin{equation}
    Z_{DF} =  \Biggl \langle  e^{\alpha_1 \phi (0)} \ e^{\alpha_2 \phi (x)} \ e^{\alpha_3 \phi (1)}  \ e^{\alpha_4 \phi (\infty)}  \left( \int_0^q e^{b \phi (z)} \right)^{N_1} \left( \int_0^1 e^{b \phi (z)} \right)^{N_2}\Biggr \rangle  =
    \label{18}
\end{equation}
  \begin{equation}
 = q^{\frac{\alpha_1 \alpha_2}{2 \beta}}  (1-x)^{\frac{\alpha_2 \alpha_3}{2 \beta}} \prod_{i=1}^{N_1} \int_0^q dz_i \prod_{i=N_1 + 1}^{N_1+N_2} \int_0^1 dz_i  \prod_{i \leq j} (z_j - z_i) ^{2 \beta} \prod_i z_i^{\alpha} (z_i - q)^{\alpha_2} (z_i - 1)^{\alpha_3}
 \label{19}
  \end{equation}
Excluding the pre-factor, we can write this partition function   as
\begin{equation}
    Z_{DF} = \int \prod_{i=1}^{N} dz_i  \prod_{i} (z_j - z_i)^{2 \beta} \prod_i z_i^{\alpha} (z_i - q)^{\alpha_2} (z_i - 1)^{\alpha_3}
=  \int \prod_{i=1}^{N} dz_i  \prod_{i} (z_j - z_i)^{2 \beta} e^{- \frac{1}{g_s} \sum_i V(z_i)}
    \label{21}
\end{equation}
where $g_s$ is a constant and $V(z)$ is the potential
\begin{equation}
    V(z_i) = -\alpha_1\ ln z - \alpha_2 \ ln (z-q) - \alpha_3 \ ln (z-1)
\end{equation}
This formula is already typical for eigenvalue matrix models \cite{UFN3,Mir}, in the particular case of (\ref{21}) it is of Penner-type \cite{Penner}. It is widely used in topological string theory \cite{DV}, superconformal gauge theory, and in the study \cite{MMS,MMS2}  of  AGT relations \cite{AGT}.\\
\newline
In the case of WZW model the integrals like (\ref{19}) possess additional symmetries and the associated matrix models deserve additional attention and further study. It was originated by Wakimoto \cite{Wak} in the case of $\mathfrak{g}=sl(2)$, continued in \cite{Morsl3} for $\mathfrak{g}=sl(3)$ and led to a general bosonization theory of WZW model and Kac-Moody alrebras in \cite{GMMOS,FeiginFrenkel}. The story begins with a polynomial expression of the Kac-Moody currents $J^a(z)$ through the $U(1)$-ones $\partial\phi^a$ and their derivatives, and for generic values of $k$ involves as many as $ dim_{\mathfrak{g}}$ free fields $\phi^a$. Actually, \cite{GMMOS} was more concentrated on bosonization of currents for generic simple algebras,
and made only the first steps towards calculating conformal blocks (though did it for generic Riemann surfaces). The study of WZW conformal blocks was largely abandoned -- together with that of generic conformal theories. Thus neither the correlators nor matrix models with additional symmetries remain underinvestigated -- what is neither just, nor convenient for practical purposes. The present paper is a step towards curing this situation.\\
\newline
Alternatively, as we mentioned above, conformal blocks of WZW model can be found by solving the Knizhnik--Zamolodchikov (KZ) equation \cite{KZ},
which is a system of partial differential equations on the correlator of WZW primary fields. Such correlator can be also considered as the sum of holomorphic conformal blocks. For a correlator \( F(z_1, \dots, z_n) \)  in a WZW model based on an affine Lie algebra \( \hat{\mathfrak{g}} \) at level \( k \), the KZ equation is:
\begin{equation}
\left( (k + h^\vee) \frac{\partial}{\partial z_i} - \sum_{\substack{j = 1 \\ j \neq i}}^{n} \frac{\sum_a T^a_{(i)} \otimes T^a_{(j)} }{z_i - z_j} \right) F(z_1, \dots, z_n) = 0
\label{kz}
\end{equation}
where \( k \) is the level of the affine Kac--Moody algebra, \( h^\vee \) is the dual Coxeter number of \( \hat{\mathfrak{g}} \). For the algebra of type ($A_n, D_n,E_n$), the dual Coxeter number equals the Coxeter number. For the algebra $sl_N$,$ h^\vee =N$.   \( T^a \) are generators of \( \mathfrak{g} \) and $\sum_a T^a_{(i)} \otimes T^a_{(j)} $ acting on the \( i \)-th and \( j \)-th factors. Bosonization of WZW model implies that KZ has solutions in terms of multiple integrals of hypergeometric type, coming from the free-field representation of conformal blocks, which involve integrals over positions of screenings. While some integral solutions to KZ equations has been studied from different points of view and are known in the literature, \cite{SV, SV2, Suss, Fuchs, Tarasov, Ribault,RibaultTeshner, GHLR} their explicit relation to formulas like (\ref{19}) remains to be carefully derived. With that goal, we provide an example of such a calculation for the simplest conformal block of $\hat{sl}(3)_k$ WZW.
\bigskip

To give an overall picture of the derivation, from the bosonization process to the hypergeometric form of the conformal block, we present the following diagram below. We summarize the whole idea  and list all the steps, needed for finding a non-vanishing holomorphic correlator (conformal block) in WZW model.
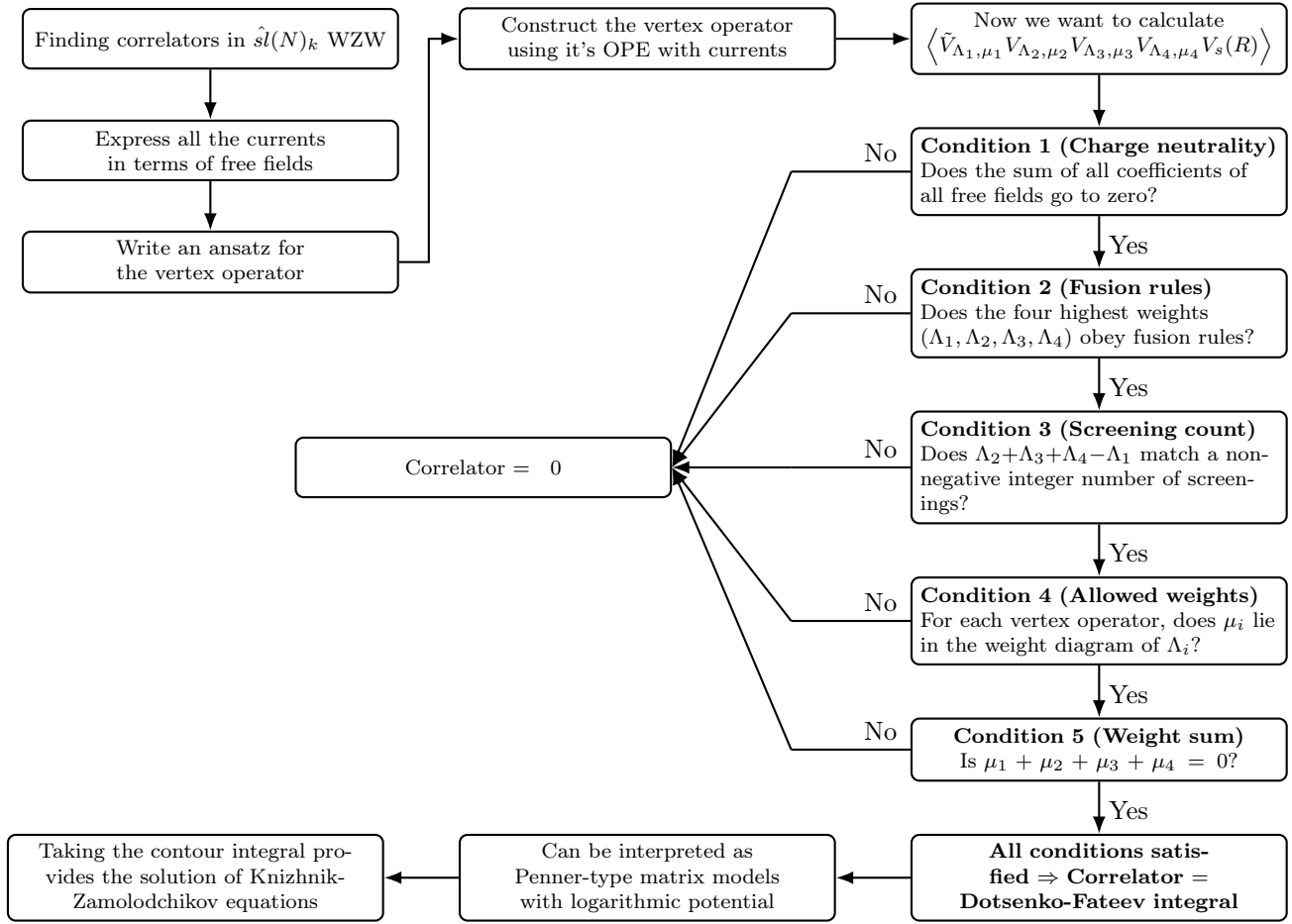
\tikzset{
  >=Latex,
  rbox/.style   = {rectangle, rounded corners=3pt, draw, thick,
                   align=center, font=\footnotesize,
                   text width=48mm, minimum height=8mm},
  cond/.style   = {rbox, align=left},
  spine/.style  = {thick,->}
}
\begin{center}
\begin{tikzpicture}[node distance=7mm]

\node[rbox, anchor=west] (L0) at (-9,2.8)
  {Finding correlators in $\hat{{sl}}(N)_k$ WZW};

\node[rbox, below=15mm of L0.west, anchor=west] (L1)
  {Express all the currents in terms of free fields};

\node[rbox, below=15mm of L1.west, anchor=west] (L11)
  {Write an ansatz for the vertex operator};

\node[rbox, right=8mm of L0.east] (Core)
  {Construct the vertex operator using it's OPE with currents};

\node[rbox, right=10mm of Core.east] (C0)
  {Now we want to calculate $\Bigr \langle \tilde{V}_{\Lambda_1,\mu_1}V_{\Lambda_2,\mu_2}V_{\Lambda_3,\mu_3}V_{\Lambda_4,\mu_4} V_s(R) \Bigr \rangle$ };

\draw[spine] (L0) -- (L1);
\draw[spine] (L1) -- (L11);
\draw[spine] (L11.east) -- ++(4mm,0) |- (Core);
\draw[spine] (Core) -- (C0);

\node[cond, below=of C0] (C1)
  {\textbf{Condition 1 (Charge neutrality)} \\ Does the sum of all coefficients of all free fields go to zero? };

\node[cond, below=of C1] (C2)
  {\textbf{Condition 2 (Fusion rules)}\\Does the four highest weights $(\Lambda_1,\Lambda_2,\Lambda_3,\Lambda_4)$ obey fusion rules?};

\node[cond, below=of C2] (C3)
  { \textbf{Condition 3 (Screening count)}\\Does $\Lambda_2{+}\Lambda_3{+}\Lambda_4{-}\Lambda_1$ match a non-negative integer number of screenings?};

\node[cond, below=of C3] (C4)
  {\textbf{Condition 4 (Allowed weights)}\\For each vertex operator, does $\mu_i$ lie in the weight diagram of $\Lambda_i$?};

\node[rbox, below=of C4] (C5)
  {\textbf{Condition 5 (Weight sum)}\\ Is $\mu_1+\mu_2+\mu_3+\mu_4=0$?};

\node[rbox, below=of C5] (Success)
  {\textbf{All conditions satisfied $\Rightarrow$  Correlator $=$ Dotsenko-Fateev integral}};

\coordinate (BusTop)    at ($(C0.west)+(-16mm,6mm)$);
\coordinate (BusC1)     at ($(C1.west)+(-16mm,0)$);
\coordinate (BusC2)     at ($(C2.west)+(-16mm,0)$);
\coordinate (BusC3)     at ($(C3.west)+(-16mm,0)$);
\coordinate (BusC4)     at ($(C4.west)+(-16mm,0)$);
\coordinate (BusC5)     at ($(C5.west)+(-16mm,0)$);
\coordinate (BusBottom) at ($(C5.west)+(-16mm,6mm)$);

\node[rbox, left=32mm of C3.west] (Fail)
  { Correlator $=0$};

\draw[thick] (C1.west) -- ++(-8mm,0) node[pos=.5, above]{No} -- (BusC1);
\draw[thick] (C2.west) -- ++(-8mm,0) node[pos=.5, above]{No} -- (BusC2);
\draw[thick] (C3.west) -- ++(-8mm,0) node[pos=.5, above]{No} -- (BusC3);
\draw[thick] (C4.west) -- ++(-8mm,0) node[pos=.5, above]{No} -- (BusC4);
\draw[thick] (C5.west) -- ++(-8mm,0) node[pos=.5, above]{No} -- (BusC5);

\draw[spine] (BusC1) -- (Fail.east);
\draw[spine] (BusC2) -- (Fail.east);
\draw[spine] (BusC3) -- (Fail.east);
\draw[spine] (BusC4) -- (Fail.east);
\draw[spine] (BusC5) -- (Fail.east);

\node[rbox, left=10mm of Success] (Lresult)
  {Can be interpreted as Penner-type matrix models with logarithmic potential};
\node[rbox, left=10mm of Lresult] (Presult)
  {Taking the contour integral provides the solution of Knizhnik-Zamolodchikov equations};

\draw[spine] (C0) -- (C1);
\draw[spine] (C1) -- node[pos=.55, right]{Yes} (C2);
\draw[spine] (C2) -- node[pos=.55, right]{Yes} (C3);
\draw[spine] (C3) -- node[pos=.55, right]{Yes} (C4);
\draw[spine] (C4) -- node[pos=.55, right]{Yes} (C5);
\draw[spine] (C5) -- node[pos=.55, right]{Yes} (Success);
\draw[spine] (Success.west) -- (Lresult.east);
\draw[spine] (Lresult) -- (Presult);

\end{tikzpicture}
\end{center}
Throughout the paper, we will follow this path for particular algebras and hopefully, this problem will be appreciated and the work will be extended to more general settings.\\

While it is technically difficult to calculate multipoint correlators by solving the KZ equation, the free-field representation creates room for further development of the global problem, which may be more practical. Our work provides strong intuition about a larger class of open problems in multipoint WZW theories for future study, which we believe to be very promising. Establishing a connection between the WZW model of arbitrary affine algebra, its representations of any type in the case of multipoint correlators, and developing a general method for the explicit derivation of contour integral representations in this broad context, is our main goal for the future. The questions of analyzing the dimension of the solution space of the KZ equation and finding a concrete technique to extract generalized hypergeometric functions from such integrals arise naturally along these lines.\\

This paper is organized as follows. In section 2, we demonstrate the bosonization process of $\hat{\mathfrak{sl}}(2)_k$ WZW model
and explicitly calculate the two point, three point and four point correlator of the vertex operators. We also show the process of getting the DF integral in detail. Then we apply the results to the case of different spin representations and study their global and local $SU(2)$ symmetry. In section 3, we repeat this process of DF integral derivation for $\hat{\mathfrak{sl}}(3)_k$. In section 4, we illustrate the bigger picture of WZW conformal blocks from the general point of view. Since we consider only holomorphic (meromorphic) conformal blocks in this paper, the work ``correlator'' will actually mean conformal block. Fateev-Dotsenko formalism can be modified to provide true correlators (modular invariant bilinears of conformal blocks), but we will not discuss them in the present text.

\section{Free field realization of $\hat{\mathfrak{sl}}(2)_k$ WZW model }

\subsection{Wakimoto's bosonization of $\hat{\mathfrak{sl}}(2)_k$ WZW model }
For a first order or bosonic ghost system of $(1,0)$, which is a pair of bosonic fields $W(z)$ and $\chi(z)$ with conformal weight of $1$ and $0$, we can write   the bosonized currents of the affine $\hat{\mathfrak{sl}}(2)_k$  algebra at level \( k \), with \( q = \sqrt{k + 2} \), which are given by:

\begin{align}
J_+(z) &= \frac{i}{\sqrt{2}}\, W(z) \\[1ex]
H(z) &= \frac{iq}{\sqrt{2}}\, \partial \phi(z) + iW(z)\chi(z) \\[1ex]
J_-(z) &= \frac{i}{\sqrt{2}} \left[ W(z)\chi^2(z) - i \sqrt{2}q\, \chi(z)\partial \phi(z) - q^2 \partial \chi(z) \right]
\end{align}
where $\phi(z)$ is a free scaler field or free boson. Using the Sugawara construction, we can obtain the energy-momentum tensor
\begin{equation}
    T = \frac{1}{q^2} : J_+J_- + J_-J_++H^2:
\end{equation}

\subsection{Second-level bosonization}

Now, to express the $W-\chi$ bosonic system fully in terms of free scalar fields, we need to further bosonize this $W-\chi$ system itself. Namely, we introduce two free bosons $u(z)$, $v(z)$ and together with $\phi (z)$, we can rewrite the currents of \( \hat{\mathfrak{sl}}(2)_k \) completely in terms of these three scaler fields. For this, the $W$ and $\chi$ take the following form
\begin{equation}
\begin{aligned}
    W =& - \partial \xi e^{-u} = -i \partial v e^{-u+iv} \ \ \text{and}  \ \ \chi = \eta e^u = e^{u-iv}; \\& \text{where } \ \ \xi (z) = e^{iv(z)} \ \ \text{and} \ \  \eta (z) = e^{-iv(z)}
    \end{aligned}
 \end{equation}
where the OPE between $\xi (z)$ and $\eta (z)$ is
\begin{equation}
 \xi (z) \eta (z') = \frac{1}{z-z'} + \dots
\end{equation}
and the OPE of the scaler fields $u(z)$ and $v(z)$ are:
\begin{equation}
\begin{aligned}
   & u(z) u(z') = - \text{log} (z-z') + \dots \\& v(z) v(z') = - \text{log} (z-z') + \dots \\& u(z) v(z') = 0
\end{aligned}
\end{equation}
now we can rewrite the currents fully in terms of these three scaler fields $u,v \ \text{and} \ \phi $ as
\begin{align}
J_+(z) &= \frac{1}{\sqrt{2}}\, \partial v(z)\, e^{-u(z) + i v(z)} \\[1em]
H(z) &= \frac{iq}{\sqrt{2}}\, \partial \phi(z) + \partial v(z) \\[1em]
J_-(z) &= \frac{1}{\sqrt{2}} \left[ \sqrt{2} q\, \partial \phi(z) - i q^2\, \partial u(z) + \left(1 - q^2\right) \partial v(z) \right] e^{u(z) - i v(z)}
\label{266}
\end{align}
Again using the Sugawara construction, we can also get the expression for the energy-momentum tensor as
\begin{equation}
T(z) = \frac{1}{2q^2} \left( 2 J_+(z) J_-(z) + H^2(z) \right)
= T_u(z) + T_v(z) + T_{\phi}(z)
\end{equation}
where
\begin{equation}
    T_{\zeta} = \frac{1}{2}(\partial \zeta)^2 + i \sqrt{2} \alpha_{0,\zeta} \partial^2 \zeta; \ \ \ \ \ \zeta = u,v,\phi
\end{equation}
with
\begin{equation*}
    \alpha_{0,u} =  \alpha_{0,v} = \frac{i}{2\sqrt{2}}; \ \ \text{and} \ \  \alpha_{0,\phi} = -\frac{1}{2q}
\end{equation*}

\subsection{Vertex operators}

We have the generators in terms of free fields and now, we want to construct the primary vertex operators $V_{j,m}$ for the algebra $\hat{\mathfrak{sl}}(2)_k$ also in terms of the free fields $\phi, u, v$. For this, we consider the following ansatz for the general vertex operator as follows:
\begin{equation}
    V = \exp(i \lambda \phi + \sigma u + i \eta v)
    \label{299}
\end{equation}
where $\lambda, \sigma$ and $\eta$ are three parameters need to be determined. Our ansatz (\ref{299}) has some conditions due to the OPE with the currents. From these conditions, we can find these parameters and write the final form of the vertex operator. The ansatz (\ref{299}), must satisfy
\begin{itemize}
    \item The OPE of $V$ with $J_{\pm} (z)$ should have a pole of first order.
    \item The OPE of $V$ with $H(z)$ gives the residue of the weight $m$.
\end{itemize}
The OPE of $V$ with the currents $J_+$, $J_-$, $H$:
\begin{equation}
\begin{aligned}
H(z) V (\omega) &=\frac{ \left( - \frac{q \lambda}{\sqrt{2}} + i \eta \right) }{(z-\omega)} V(\omega) + \cdots \\
J_+(z) V (\omega) &= \frac{\eta}{(z-\omega)^{1 - \sigma - \eta}} \frac{i}{\sqrt{2}} V_{\text{raised}} (\omega ) + \cdots \\
J_-(z) V(\omega) &= \frac{1}{(z-\omega)^{1 + \eta + \sigma}} \frac{i}{\sqrt{2}}
\left[ \sqrt{2} q \lambda - q^2 \sigma + \eta (1 - q^2) \right] V_{\text{lowered}}(\omega) + \cdots
\end{aligned}
\end{equation}

Now, we have several simple steps to find the general formula for the operator $V$. First of all, to fulfill the first condition, we should have
\begin{equation}
    1  - \sigma - \eta = 1 \ \Rightarrow  \  \eta = -\sigma
\end{equation}
For the highest weight state, we require that the highest weight vector $V_j$ satisfies the following relation
\begin{equation}
    J_+ \bra{V_j} =0
\end{equation}
where this $j$ can be interpreted as the spin. Now, for this relation to hold, we have to put $\eta = 0$. This also gives $\sigma = 0$. So we get the vertex operator  for the highest weight state
\begin{equation}
    V_j = e^{i\lambda \phi(z)}
\end{equation}
To find $\lambda$, we use the second condition. For the highest weight state $m=j$, we get
\begin{equation}
    \frac{q\lambda}{\sqrt{2}} +\sigma = j  \ \Rightarrow  \ \lambda = \frac{j \sqrt{2}}{q}
\end{equation}
where $m$ can be interpreted as the magnetic number. So, the highest weight operator takes the following form
\begin{equation}
    V_j(z) = \exp \left( i  \frac{j \sqrt{2}}{q}\phi(z)\right)
\end{equation}
Now, by acting the lowering operator, we can construct the vertex operator of any state. For example, to get the vertex operator for $\left( \frac{1}{2}, -\frac{1}{2}\right)$, we act the lowering operator on the highest weight vector. For higher spin $j$, we can act repeatedly with the lowering operator, according to the allowed values $m=j,j-1,j-2,\dots -j$.
\begin{equation}
\begin{aligned}
    & J_-(w) V_{j,j}(z) = \frac{i \sqrt{2}j }{w-z}V_{j,j-1}(z),   \ \  \ \ \ \ \ \ \ \ \ \ \ \ \ \ \ \  V_{j,j-1} = \chi V_{j,j} (z)
    \\& J_-(w) V_{j,j-1} (z) = \frac{i \sqrt{2} \left( j-\frac{1}{2} \right) }{w-z} V_{j,j-2}(z), \ \ \ \ \ \ \  V_{j,j-2} =  \chi^2 V_{j,j}(z)
    \\& J_-(w) V_{j,j-2} (z) = \frac{i \sqrt{2}(j-1)}{w-z} V_{j,j-3}(z), \ \ \ \  \ \ \ \ \ V_{j,j-3} =  \chi^3 V_{j,j}(z)
    \\& \vdots
    \\& J_-(w)V_{j,j-k}(z) =
\frac{i\sqrt{2}\left(j-\frac{k}{2}\right)}{w-z}\,V_{j,j-k-1}(z),  \ \ \  V_{j,j-k}(z)=\chi^k(z)V_{j,j}(z),
    \\& \vdots
    \\& J_-(w)V_{j,-j+1}(z) = \frac{i}{\sqrt{2}}\frac{1}{w-z}\,V_{j,-j}(z), \ \ \  \ \  \ \ \ \ \ V_{j,-j}(z)=\chi^{2j}(z)V_{j,j}(z)
    \\& J_-(w)V_{j,-j}(z) = 0  \\&
\end{aligned}
\end{equation}
 Using $m=j-k$, we can write the general formula for the vertex operators in terms of  $j$ and $m$ as
\begin{equation}
    V_{j,m } = \chi^{j-m} \ {\exp} \left( i\frac{\sqrt{2} j}{q} \phi \right) = \exp \left( i\frac{\sqrt{2} j}{q} \phi + (j-m) (u-iv) \right), \ \ \ -j \leq m \leq j 
    \label{288}
\end{equation}
The conformal dimension of this operator depends on the spin$-j$ and will have the following form
\begin{equation}
    \Delta_j = \frac{j (j+1)}{q^2} = \frac{j(j+1)}{k+2}
    \label{2990}
\end{equation}
\subsection{Correlator of Vertex Operators}

\subsubsection{Charge neutrality condition and screening operator}
To calculate the correlator, which are holomorphic conformal blocks or their linear combination, we reduced this problem to the calculation of the correlators of these vertex operators (\ref{288}), which contain three scaler fields $\phi$, $u$ and $v$. The central charge of the WZW model is non-zero, which means that it suffers from a gravitational or conformal anomaly. If we have a conformal theory with zero central charge, then the correlator will depend only on the insert point of the vertex operators. But in case of WZW model on a sphere, the correlator also depends not only on the operator insertion points but also on the metric of the sphere as well. Therefore, to calculate the correlator here, we need to insert another operator at the point of singularity ($R$) of the metric. If we take the metric as $ds^2 = dz d\bar{z}$, then the location of $R$ will be at $\infty$. This particular operator is called the vacuum charge. Therefore, in the correlator, we need to insert the vacuum charge of the following type
\begin{equation}
    V_s(R) = \exp \left( \frac{i\sqrt{2} }{q} \phi (R) + u(R) - iv(R) \right)
\end{equation}
Before we proceed with the calculation of the correlator, let us take a moment to explore the charge neutrality condition which we need to fulfill to calculate a non-zero correlator. For this, let us consider the theory of the free massless scalar field $S = \frac{1}{8\pi} \int d^2 x (\partial_\mu \phi)^2$. Under the global shift symmetry $\phi(x) \to \phi(x) + a, \ a \in \mathbb{R}$, the action is invariant: $S[\phi(x)+a ] = S[\phi(x)]$ . \\
\newline
Now, using the path integral, we want to find the correlation function of the following vertex operators with the charge $  \alpha_k$
\begin{equation}
      \Biggl \langle \prod_{k=1}^n e^{i \alpha_k \phi(x_k)}  \Biggl \rangle = \int \mathcal{D} \phi \prod_{k=1}^n e^{i \alpha_k \phi(x_k)} \ e^{-S[\phi]}
      \label{300}
\end{equation}
For the theory to be invariant under $\phi \to \phi+a$, we must have
\begin{equation}
    \delta \Biggl \langle \prod_{k=1}^n e^{i \alpha_k \phi(x_k)}  \Biggl \rangle = 0
    \label{311}
\end{equation}
To calculate that, we can look at how the vertex operator transforms
\begin{equation}
    \prod_{k=1}^n e^{i \alpha_k (\phi (x)+a)} = \prod_{k=1}^n e^{i \alpha_k\phi (x)} e^{i \alpha_k a} = \left( \prod_{k=1}^n e^{i \alpha_k \phi (x)} \right) e^{i a \sum_{k=1}^n \alpha_k}
\end{equation}
Then the correlator that corresponds to the shift will be
\begin{equation}
   \Biggl \langle \prod_{k=1}^n e^{i \alpha_k ( \phi(x_k) + a)}  \Biggl \rangle =  \int \mathcal{D} (\phi+a) \prod_{k=1}^n e^{i \alpha_k \phi(x_k)} e^{i a \sum_{k=1}^n \alpha_k} \ e^{-S[\phi +a ]} = e^{i a \sum_{k=1}^n \alpha_k}  \Biggl \langle \prod_{k=1}^n e^{i \alpha_k \phi(x_k)}  \Biggl \rangle
\end{equation}
Then (\ref{311}) will be
\begin{align}
    \delta \Biggl \langle \prod_{k=1}^n e^{i \alpha_k \phi(x_k)}  \Biggl \rangle  &= \Biggl \langle \prod_{k=1}^n e^{i \alpha_k ( \phi(x_k) + a)}  \Biggl \rangle  - \Biggl \langle \prod_{k=1}^n e^{i \alpha_k \phi(x_k)}  \Biggl \rangle = 0 \\ &= \left( e^{i a \sum_{k=1}^n \alpha_k} -1 \right) \Biggl \langle \prod_{k=1}^n e^{i \alpha_k \phi(x_k)}  \Biggl \rangle = 0
\end{align}
For an infinitesimal $a$, expand the exponential to the first order
\begin{equation}
    e^{i a \sum_{k=1}^n \alpha_k} = 1 + i a \sum_{k=1}^n \alpha_k + \mathcal{O} (a^2)
\end{equation}
Then finally, we have
\begin{equation}
   \left( i a \sum_{k=1}^n \alpha_k \right) \Biggl \langle \prod_{k=1}^n e^{i \alpha_k \phi(x_k)}  \Biggl \rangle = 0
   \label{377}
\end{equation}
The expression (\ref{377}) basically says, if we want to get a non-zero correlator, the sum of all the charges should be zero.
\begin{equation}
    \boxed{ \sum_{k=1}^n \alpha_k =0}
\end{equation}
This is known as the charge neutrality condition. In our case, if we look at the vertex operators (\ref{288}), their charge neutrality condition actually depends on the spin representation-$j$. Now, if one wants to calculate such a correlator with the vacuum charge, the charge neutrality condition does not hold, and we end up with a vanishing correlator. To make the correlator non-vanishing, we need to somehow follow the charge neutrality condition. In this case, one idea is to insert another operator into the correlator such that the total charge vanishes. This kind of operator is known as the Feigin-Fuks operator \cite{FF} $Q$ ( also called the screening operator or the screening charge). For $\hat{sl}(2)_k$, which has been defined with the following current $J(t)$:
\begin{equation}
    Q = \oint J(t) = \oint \exp \left( - \frac{i \sqrt{2}}{ q} \phi(t) - u(t) +iv(t) \right) \frac{1}{i} \partial v(t)
\end{equation}
Now, we can proceed to the calculation of the two point function. One can see that only the correlator of these operators (\ref{288}) will not fulfill the charge neutrality condition, even for any $j$. At this moment, we need to introduce a dual/reflected operator of (\ref{288}). To preserve conformal invariance, we need that the reflected operator has the same conformal dimension (\ref{299}) as that in (\ref{288}). Namely, we introduce the following operator:
\begin{equation}
    \tilde{V}_{j, \ m'-j} = V_{-1-j, \ 1+j+m'}
    \label{499}
\end{equation}
The goal of this operator is clear. This will provide us with a different combination of free fields in the exponential of vertex operators, which will help us to cancel out the total charge of all operators and finally fulfill the charge neutrality condition.

\subsubsection{Two-point function}
So, our stage is ready with all necessary ingredients, and we can now head towards the calculation of a simple two-point correlator.
$$
 \braket{V_{j, \ m-j }(z) V_{-1-j, \ 1+j+m'}(0) V_s (R)} =
$$
\begin{equation}
\Biggl \langle \exp{\left( i\frac{\sqrt{2} j}{q} \phi + (j-m) (u-iv) \right)} \exp{\left( i\frac{\sqrt{2} (-1-j)}{q} \phi + (-1-j-m') (u-iv) \right)} \exp{\left( \frac{i\sqrt{2} }{q} \phi (R) + u(R) - iv(R) \right)} \Biggl \rangle
    \label{411}
\end{equation}
We see that the total charge is zero even without inserting the screening charges. Therefore, the two-point function is the only simplest case where without inserting the screening charge we can still get the non-vanishing correlator. Using the OPE between the fields, one can easily calculate (\ref{411}) and it will be
\begin{equation}
    = \  z^{-2\Delta_j} \delta_{m+m',0}
\end{equation}
where $\Delta_j$ is the conformal dimension of the operators (\ref{288}). Now, we can calculate the correlator for different levels $k$ for $\hat{sl}(2)_k$. Table (\ref{T1}) above shows several two point functions for different $k$. Note that for every level, we will have $j=0$, which corresponds to $m=0$ and gives the value of correlator $1$. For convenience, we exclude the part of $j=0$ from each level in the table.

\begin{table}[H]
		\centering
		\begin{tabular}[h!]{c|c|c|c|c}
			$k$ & $j$ & $\{m,m'\}$ & $\Delta_j$ & $ \braket{V_{j, \ m-j }(z)\tilde{V}_{j, \ m'-j} V_s (R)} $   \\ &&&\\
			\hline &&&\\
            $k=1$ & $j=\frac{1}{2}$ & $ \{ \frac{1}{2}, -\frac{1}{2}\}$ & $ \frac{1}{4}$ & $ z^{-\frac{1}{2}}$\\&&&\\
			\hline &&&\\
            $k=2$ & $j=\frac{1}{2}, 1$ & $ \{ \frac{1}{2}, -\frac{1}{2}\}, \{1,-1 \}$ & $ \frac{3}{16}, \frac{1}{2}$ & $z^{-\frac{3}{8}},z^{-1}$   \\&&&\\
            \hline &&&\\
              $k=3$ & $j=\frac{1}{2}, 1,  \frac{3}{2}$ & $ \{ \frac{1}{2}, -\frac{1}{2}\}, \{1,-1 \},  \{ \frac{3}{2}, -\frac{3}{2}\}$ & $\frac{3}{20}, \frac{2}{5}, \frac{3}{4}$ & $z^{-\frac{3}{10}},z^{-\frac{4}{5}}, z^{-\frac{3}{2}}$   \\&&&\\
                   \hline &&&\\
              $k=4$ & $j=\frac{1}{2}, 1,  \frac{3}{2},2$ & $ \{ \frac{1}{2}, -\frac{1}{2}\}, \{1,-1 \},  \{ \frac{3}{2}, -\frac{3}{2}\}, \{ 2,-2\}$ & $ \frac{1}{8}, \frac{1}{3}, \frac{15}{24}, 1$ & $z^{-\frac{1}{4}},z^{-\frac{2}{3}}, z^{- \frac{15}{12}}, z^{- \frac{1}{2}}$   \\&&&\\
              \hline &&&\\
              $k=5$ & $j= \frac{1}{2}, 1,  \frac{3}{2},2,\frac{5}{2}$ & $ \{ \frac{1}{2}, -\frac{1}{2}\}, \{1,-1 \},  \{ \frac{3}{2}, -\frac{3}{2}\}, \{ 2,-2\}, \{\frac{5}{2}, -\frac{5}{2} \}$ & $\frac{3}{28}, \frac{2}{7}, \frac{15}{28}, \frac{6}{7}, \frac{5}{4}$  & $z^{-\frac{3}{14}},z^{-\frac{4}{7}}, z^{ - \frac{15}{14}}, z^{- \frac{12}{7}}, z^{-\frac{5}{2}}$   \\&&&\\
		\end{tabular}
  \caption{ Two point correlators of vertex operators in $\hat{sl}(2)_k$ for different level }
  \label{T1}
	\end{table}
\subsubsection{Three-point function}
So far we have constructed two operators that depend on $j$ and $m$. For convenience in writing, we denote these operators as follows
\begin{align}
V_{j,m} =  V_{j, \ m-j} &= \text{exp} \left( i\frac{\sqrt{2} j}{q} \phi + (j-m) (u-iv) \right), &\\
\tilde{V}_{j,m} =  V_{-1-j, \ 1+j+m} &= \exp{\left( i\frac{\sqrt{2} (-1-j)}{q} \phi + (-1-j-m) (u-iv) \right)}
\end{align}
From now on, for all correlators of $\hat{sl}(2)_k$ WZW, these two operators will be the main ingredient in the derivation of the DF integral. As these operators depend on both $j$ and $m$, so, for different levels, we can construct different vertex operators. Let us look at an example of this.\\
\newline
Let us first fix three points $z_1 = 0, \ z_2 = 1, \ z_3 = \infty$. Then we need to put two operators on these points. As a simple example, we consider the case of fundamental representation when $j=\frac{1}{2}$. For this, we have $m=\{ \frac{1}{2}, -\frac{1}{2}\} $. So we can actually construct four operators for the same $j=\frac{1}{2}$ and these $m$. Let us write them down
\begin{align}
    V_{\frac{1}{2}; \frac{1}{2}} &= \exp\left( \frac{i}{\sqrt{2}q} \, \phi \right), &
     V_{\frac{1}{2};-\frac{1}{2}} &= \exp\left( \frac{i}{\sqrt{2}q} \, \phi + u - iv \right)
     \label{544}
\end{align}
The reflected operators will be
\begin{align}
\tilde{V}_{\frac{1}{2}; \frac{1}{2}} &= \exp\left( -\frac{3i}{\sqrt{2}q} \, \phi - 2u + 2iv \right), &
\tilde{V}_{\frac{1}{2}; -\frac{1}{2}}^- &= \exp\left( -\frac{3i}{\sqrt{2}q} \, \phi - u + iv \right)
\label{545}
\end{align}
Now, for a three-point correlator, we need to choose three operators from these four. Note that, for the sake of charge neutrality condition, we need to take one reflected operator from these two. In case of four-point correlator (which we explain in the next subsection), we must have to take any of these reflected operators as well, and the other three operators will be without reflection (\ref{544}). So we can start trying to calculate the following correlator
\begin{equation}
   \Bigr \langle \tilde{V}_{\frac{1}{2}; -\frac{1}{2}} (0) \  V_{\frac{1}{2}; \frac{1}{2}} (1) \ V_{\frac{1}{2}; -\frac{1}{2}} (\infty) V_s(R) \Bigr \rangle
   \label{655}
\end{equation}
In case of a two-point correlator, to get a non-zero answer, we have seen that $\sum_i m_i = 0$ as we had only two operators with the same $m$ in opposite sign. In that case, no screening charges were needed to satisfy the charge neutrality condition. It was automatically held for two operators and vacuum charge. So, towards a non-zero correlator, the first condition we need to check for the operators in the correlators is
\begin{equation}
    \sum_i m_i = 0
    \label{577}
\end{equation}
If this condition holds, we can proceed to impose the charge neutrality condition. This requires that, for each field at the different insertion points, the total charge vanishes. However, we observe that in (\ref{655}) the condition (\ref{577}) is not satisfied. If we change $j$ from the fundamental representation to a higher representation, then it may be possible to choose different values of $m$ so that (\ref{577}) holds. We will address this issue in the next subsections and construct a non-vanishing correlator. For the moment, let us instead consider the four-point correlator with all $j=\frac{1}{2}$.

\subsubsection{Four-point function}
For the four insertion points $z_1=0, \ z_2=1, \ z_3 = x, \ z_4 = \infty$, we now compute the four-point correlator for the given operators.(\ref{544}), (\ref{545}):
\begin{equation}
     \Bigr \langle \tilde{V}_{\frac{1}{2}; -\frac{1}{2}} (0) \  V_{\frac{1}{2}; \frac{1}{2}} (1) \ V_{\frac{1}{2}; \frac{1}{2}} (x) \  V_{\frac{1}{2}; -\frac{1}{2}} (\infty) V_s(R) \Bigr \rangle
\end{equation}
We see that (\ref{577}) holds and now we need to check the charge neutrality condition. If we look at the fields $\phi, u, v$ at our desired points \\
\newline
at $z_1=0$:
\begin{equation}
    \tilde{V}_{\frac{1}{2}; -\frac{1}{2}}(0) = \exp\left( -\frac{3i}{\sqrt{2}q} \, \phi(0) - u(0) + iv(0) \right)
\end{equation}
at $z_2 = 1$:
\begin{equation}
V_{\frac{1}{2}; \frac{1}{2}} (1)= \exp\left( \frac{i}{\sqrt{2}q} \phi(1) \, \right)
\end{equation}
at $z_3 = x$:
\begin{equation}
V_{\frac{1}{2}; \frac{1}{2}} (x)= \exp\left( \frac{i}{\sqrt{2}q} \phi(x) \, \right)
\end{equation}
at $z_4 = \infty$:
\begin{equation}
 V_{\frac{1}{2}; -\frac{1}{2}} (\infty) = \exp\left( \frac{i}{\sqrt{2}q} \, \phi(\infty) + u(\infty) - iv(\infty) \right)
\end{equation}
at $R=\infty$:
\begin{equation}
     V_s(\infty) = \exp \left( \frac{i \sqrt{2}}{ q} \phi (\infty) + u(\infty) - iv(\infty) \right)
\end{equation}
Now, let us sum the exponential for each field\\
\newline
for $\phi$
\begin{equation}
 -\frac{3i}{\sqrt{2}q} + \frac{i}{\sqrt{2}q}+ \frac{i}{\sqrt{2}q}+ \frac{i}{\sqrt{2}q} + \frac{i \sqrt{2}}{ q} \neq 0
\end{equation}
for $u$
\begin{equation}
    -1 +0+0+1 +1 \neq 0
\end{equation}
for $v$
\begin{equation}
    i +0 +0-i - i \neq 0
\end{equation}
Thus, the total charge neutrality condition is not satisfied in this case, which means we must insert screening operators. Now, if we insert the following screening operator and recompute the charges at each field, we find that the neutrality condition is indeed satisfied. This implies that we can obtain a non-vanishing correlator in this setup.
\begin{equation}
  F =    \Bigl \langle \tilde{V}_{-}(0) V_{+} (x) V_{+} (1) V_{-}(\infty)   V_s(\infty) \oint J (t) \Bigr \rangle  = \oint dt \Bigl \langle \tilde{V}_{-}(0) V_{+} (x) V_{+} (1) V_{-}(\infty)   V_s(\infty) J (t) \Bigr \rangle
     \label{39}
\end{equation}
By putting all the exponentials,
\begin{equation}
    = \oint \Biggr \langle e^{\left(-\frac{3i}{\sqrt{2}q} \, \phi(0) - u(0) + iv(0) \right)} e^{\left( \frac{i}{\sqrt{2}q} \, \phi(x) \right)} e^{\left( \frac{i}{\sqrt{2}q} \phi(1) \, \right)} e^{\left( (\frac{i}{\sqrt{2}q} + \frac{i}{ q} ) \, \phi(\infty) + 2u(\infty) - 2iv(\infty) \right)} e^{\left( - \frac{i \sqrt{2}}{ q} \phi(t) - u(t) +iv(t) \right)} \frac{1}{i} \partial v(t) \Biggr \rangle
    \label{688}
\end{equation}
We see that there are two types of contractions here. One arises from the exponential term, while the other comes from the derivative term outside the exponential, which contracts with the $v$-fields at other points. Let us first analyze the contractions coming from the exponential. We will describe the general pattern of these contractions and then consider a specific example of such a calculation.(\ref{688})\\
\newline
The two-point correlators of the free fields are given by
\begin{align*}
    \braket{ \phi(z) \phi(\omega)} &= - \log (z-\omega)\\
    \braket{u(z) u(\omega) } &= - \log (z-\omega)\\
    \braket{ v(z) v(\omega)} &= - \log (z-\omega)
\end{align*}
In a general case, we can rewrite the exponential of the correlator as follows
\begin{align}
& \Biggl \langle \exp \left( \sum_i a_i \phi(z_i) + b_i u (z_i) + c_i v (z_i) \right) \Biggl \rangle \ = \ \exp \left( \frac{1}{2} \sum_{i<j} (a_i a_j + b_i b_j + c_i c_j) \braket{\phi (z_i) \phi (z_j)} + \dots \right)
\end{align}
Since the fields are independent, we can write:
\begin{align}
    =& \ \exp \left( - \frac{1}{2} \sum_{i<j} (a_i a_j + b_i b_j + c_i c_j) \log (z_j - z_j) \right)   = \prod_{i<j} (z_i - z_j)^{ - \frac{1}{2} (a_i a_j + b_i b_j + c_i c_j)  }
\end{align}
Now, for the field $\phi$,
\begin{equation}
   \exp \left( \sum_{i<j} a_i a_j \braket{\phi (z_i) \phi(z_j)} \right) = \exp \left(- \sum_{i<j} a_i a_j \ \log (z_i - z_j) \right) = \prod_{i<j} (z_i - z_j)^{- a_i a_j}
\end{equation}
We can also obtain a similar expression for $u$ and $v$. So, the contractions in the entire exponential part give the following.
\begin{equation}
    \prod_{i<j} (z_i - z_j)^{- a_i a_j} \cdot \prod_{i<j} (z_i - z_j)^{- b_i b_j} \cdot \prod_{i<j} (z_i - z_j)^{- c_i c_j} = \prod_{i<j} (z_i - z_j)^{- ( a_i a_j + b_i b_j + c_i c_j )}
\end{equation}
Now, we have the contractions from the derivative. The derivative $\frac{1}{i} \partial v$ must be contracted with the $v$-charges at the other points. The contraction of the derivative with a vertex operator at point $k$ (with $v$-charge $c_k$) gives
\begin{equation}
    \frac{1}{i} \left(  c_k \braket{ \partial v(t) v(z_i) } \right) = \frac{1}{i} \sum_{k=1}^4 c_k \left( - \frac{1}{t-z_k} \right)
    \label{48}
\end{equation}
Now we have the full four-point correlator with all possible contractions
\begin{equation}
    \oint dt \Bigl \langle \tilde{V}_{-}(0) V_{+} (x) V_{+} (1) V_{-}(\infty)   V_s(\infty) J (t) \Bigr \rangle = \oint dt \biggr[  \prod_{1 \leq i \leq j \leq 5} (z_i - z_j)^{- ( a_i a_j + b_i b_j + c_i c_j )}    \biggr] \biggr[ \frac{1}{i} \sum_{k=1}^4 c_k \left( - \frac{1}{t-z_k} \right)  \biggr]
    \label{49}
\end{equation}
 Now, at points $z_1 = 0, z_2 = x, z_3 = 1, z_4 = \infty,z_5 = t$, excluding the contractions with infinity, we  have the following contractions
    \begin{multline}
         (0-x)^{d_{0x}} \cdot (0-1)^{d_{01}} \cdot (0-t)^{d_{0t}} \cdot (x-0)^{d_{x0}} \cdot (x-1)^{d_{x1}} \cdot (x-t)^{d_{xt}} \cdot (1-0)^{d_{10}} \cdot (1-x)^{d_{1x}} \cdot \\ \cdot (1-t)^{d_{1t}} \cdot (t-0)^{d_{10}} \cdot (t-x)^{d_{tx}} \cdot (t-1)^{d_{t1}}
    \end{multline}
Here, by the notation $d_{ij}$ we denote $d_{ij} = - ( a_i a_j + b_i b_j + c_i c_j )$. All the contractions that are independent of $t$ actually give constants, and one can explicitly calculate them from the expression above. When we integrate over $t$, we can collect all these constant terms into the overall proportionality factor and pull them out of the integral. Let us denote this coefficient by $\mathcal{G}(q,x)$. Thus, from now on our focus will be only on
\begin{equation}
    (t-0)^{d_{t0}} \cdot  (t-x)^{d_{tx}} \cdot  (t-1)^{d_{t1}}
\end{equation}
Now, to calculate $d_{t0}, d_{tx}, d_{t1}$, we need to multiply the coefficients of all three fields at each desired points.
\begin{align*}
    d_{t0} = - \left(\left( -\frac{i \sqrt{2}}{ q}  \right)\left( -\frac{3i}{\sqrt{2} q}  \right)  + (-1) (-1) + (i) (i) \right) &= \frac{3}{q^2} \\ d_{tx} = - \left( \left( -\frac{i \sqrt{2}}{ q}  \right) \left( \frac{i}{\sqrt{2} q}  \right) + (-1) (0) + (i) (0) \right) &= - \frac{1}{q^2} \\ d_{t1} = - \left( \left( -\frac{i \sqrt{2}}{ q}  \right) \left( \frac{i}{\sqrt{2} q}  \right) + (-1) (0) + (i) (0) \right) &= - \frac{1}{q^2}
\end{align*}
So, the exponential part of (\ref{688}) gives
\begin{equation}
    t^{\frac{3}{q^2} } \cdot (t-x)^{ - \frac{1}{q^2}} \cdot (t-1)^{ - \frac{1}{q^2}}
\end{equation}
Now, to deal with the contraction involving the derivative, we expand the sum in (\ref{48}) over all four points. As the $v$-charge is zero at the points $x$ and $1$, we can exclude them. Then we have
\begin{equation}
    \frac{1}{i} \sum_{k=1}^4 c_k \left( - \frac{1}{t-z_k} \right) = - \frac{1}{i} (i) \left( - \frac{1}{t- 0} \right) + \frac{1}{i} (-2i)  \left( - \frac{1}{t- \infty} \right)
    \label{diff contraction}
\end{equation}
So we left with only one term.
\begin{equation}
     \frac{1}{i} \sum_{k=1}^4 c_k \left( - \frac{1}{t-z_k} \right) = \frac{1}{t}
\end{equation}
Finally, the correlator (\ref{688}) is
\begin{equation}
\begin{aligned}
    F = \oint dt \Bigl \langle \tilde{V}_{-}(0) V_{+} (x) V_{+} (1) V_{-}(\infty)   V_s(\infty) J (t) \Bigr \rangle = \mathcal{G}(q,x) \oint dt \left[ t^{\frac{3}{q^2} } \cdot (t-x)^{ - \frac{1}{q^2}} \cdot (t-1)^{ - \frac{1}{q^2}} \cdot \frac{1}{t} \right] & = \\ = \mathcal{G}(q,x) \oint dt \left[  t^{\frac{3-q^2}{q^2}} \cdot (t-x)^{ - \frac{1}{q^2}} \cdot (t-1)^{ - \frac{1}{q^2}} \right]
\end{aligned}
\end{equation}

now recalling $q^2 = k+2$, where $k$ is the label of the affine \( \hat{\mathfrak{sl}}_2 \),
\begin{equation}
    F =  \oint dt \Bigl \langle \tilde{V}_{-}(0) V_{+} (x) V_{+} (1) V_{-}(\infty)   V_s(\infty) J (t) \Bigr \rangle = \mathcal{G}(q,x) \oint dt \left[ t^{\frac{1-k}{k+2}} \cdot (t-x)^{ - \frac{1}{k+2}} \cdot (t-1)^{ - \frac{1}{k+2}}    \right]
     \label{60}
 \end{equation}
This is the Dotsenko–Fateev integral representation \cite{DF} of the correlators of the $\hat{sl}(2)_k$ WZW model, which is a linear combination of conformal blocks. But the most important question actually appears here. How many blocks does it contain? And this is actually related to evaluating the contour itself. In our case, its just the number of closed lines encircling/separating four points — this means we have two such options. They can be represented as integrals around two cuts — between 0 and 1 and between 1 and $\infty$. Thus, the two independent contours provide two blocks, which makes the solution space of the Knizhnik–Zamolodchikov equation (\ref{kz}) two dimensional. Namely, we can then rewrite the KZ equation as

\begin{equation}
     (k+2) \ \partial_x  \left( {F}_1 + {F}_2 \right)  - \left( \frac{ \sum_a T_1^a \otimes T_2^a}{x} + \frac{ \sum_a T_1^a \otimes T_3^a}{x-1} \right) \left( {F}_1 + {F}_2 \right) = 0
    \label{kz4}
\end{equation}
which is originally a system of first order differential equations, and then converted to a scaler equation of second order of the following type
\begin{equation}
    x(1-x) \frac{\partial^2 F}{\partial x^2} + [C - (A+B+1)x] \frac{\partial F}{\partial x} - AB F = 0
    \label{hypergeo}
\end{equation}
with some coefficients $A,B,C$. In general, the solution of such an equation is a linear combination of two Gauss hypergeometric functions ${}_2F_{1}$ with coefficients $A,B,C$. These coefficients can be determined from the integral representation, and they depend on the parameters appearing in the integrand. For this particular integral one can, in fact, rewrite it explicitly as a using standard Euler integral identities. However, in higher-rank WZW models we expect integrals with many parameters and more complicated contour structures. To handle these systematically, it is useful to understand the general strategy for extracting the corresponding hypergeometric functions from such integrals. \\
\newline
Typically, solutions of equations of the form \eqref{hypergeo} are presented as power series, and we can obtain this series representation by expanding the $x$-dependent factors in the integral. Namely, we have
\begin{equation}
    F =  \oint dt \left[  t^{\frac{3}{q^2} -1} \cdot (t-x)^{ - \frac{1}{q^2}} \cdot (t-1)^{ - \frac{1}{q^2}} \right]
\end{equation}
Let's expand the $x$-term and write using the Pochhammer symbol,
\begin{equation}
    (t-x)^{-\frac{1}{q^2}} = t^{-\frac{1}{q^2}} \sum_{n\geq0} \frac{\left(\frac{1}{q^2}\right)_n}{n!} t^{-n} x^n
\end{equation}
Then we can introduce the following power series
\begin{equation}
    F = \sum_{n \geq0} A_n x^n \ \ \ \ \text{with} \ \ \ \ A_n = \frac{\left(\frac{1}{q^2}\right)_n}{n!} \oint t^{\frac{2}{q^2} -1-n} (t-1)^{-\frac{1}{q^2}}
\end{equation}
Here, in the coefficient $A_n$ we observe a Pochhammer integral identity.
\begin{equation}
    \int_0^1 t^{\alpha -1} (1-t)^{\beta-1} = \frac{\Gamma(\alpha) \Gamma(\beta)}{\Gamma(\alpha+\beta)}
\end{equation}
Then, we can express the coefficient as
\begin{equation}
    A_n = \frac{( 1/q^2)_n}{n!} \frac{\Gamma \left( \frac{2}{q^2} - n -1 \right) \Gamma \left( 1-\frac{1}{q^2} \right)}{ \Gamma \left( 1+\frac{1}{q^2} -n\right)}
\end{equation}
After using some properties of the Gamma function $\Gamma(z-1) = \Gamma(z)/(1-z)$ and simplifying the terms, we get the following recurrence relation:
\begin{equation}
    \frac{A_{n+1}}{A_n} = \frac{\left(n+\frac{1}{q^2} \right) \left( -\frac{1}{q^2} +n \right)}{ (n+1) \left(1-\frac{2}{q^2} +n \right)}
\end{equation}
The general form of hypergeometric recurrence relation is
\begin{equation}
    \frac{A_{n+1}}{A_n} = \frac{(n+A) (n+B)}{(n+1) (n+C)}
\end{equation}
so, comparing these, we get
\begin{equation}
    A= \frac{1}{q^2}; \ \ \ B= -\frac{1}{q^2}; \ \ \ C = 1-\frac{2}{q^2}
\end{equation}
The first hypergeometric solution we get is
\begin{equation}
    F_1 = f_1(x) \ \mathcal{F}(A,B,C;x) = f_1(x) \ \mathcal{F}\left( \frac{1}{q^2}, -\frac{1}{q^2},1-\frac{2}{q^2};x \right)
\end{equation}
with some coefficient $f_1(x)$, depending on $x$ and $q$. Another hypergeometric function we get is
\begin{equation}
    F_2 = f_2(x) \ \mathcal{F} (1+B-C, 1+A-C, 2-C,x) = f_2(x) \ \mathcal{F} \left( \frac{1}{q^2}, \frac{3}{q^2}, 1+\frac{2}{q^2}; x \right)
\end{equation}
Now, we can actually derive the hypergeometric differential equation (\ref{hypergeo}) if we just know what the order of the equation should be. Analysing the contour, we find that the solution space is two-dimensional, which tells us that the order of the equation is also two-dimensional. We can write the following ansatz on the integral with unknown functions and a constant.
\begin{equation}
    F = f_1(x) \ \mathcal{F}\left( \frac{1}{q^2}, -\frac{1}{q^2},1-\frac{2}{q^2};x \right)  + f_2(x) \ \mathcal{F} \left( \frac{1}{q^2}, \frac{3}{q^2}, 1+\frac{2}{q^2}; x \right)
    \label{fil}
\end{equation}
There is another way to derive this. Where we directly derive a hypergeometric differential equation on the integral representation. Let's look at that. Let the differential equation is
\begin{equation}
    a(x) \frac{\partial^2 F}{\partial x^2} + b(x) \frac{\partial F}{\partial x} + c \ F = 0
    \label{hg2}
\end{equation}
we need to find these functions and then extract the coefficients $A,B,C$ from them. Let
\begin{equation}
    F = \oint I(x,t) dt; \ \ \ \text{where}  \ \ \ I(x,t) = t^{\frac{3}{q^2} -1} \cdot (t-x)^{ - \frac{1}{q^2}} \cdot (t-1)^{ - \frac{1}{q^2}}
\end{equation}
Now taking the derivatives over-$x$
\begin{equation}
    \frac{\partial F}{\partial x} = \oint  I(x,t) \frac{1/q^2}{t-x} dt; \ \ \ \ \frac{\partial^2 F}{\partial x^2} = \oint I(x,t) \frac{1/q^2(1/q^2+1)}{(t-x)^2}dt
\end{equation}
then putting everything back to (\ref{hg2})
\begin{equation}
    \oint  I(x,t) \underbrace{\left[ a(x) \frac{\frac{1}{q^2} \left( \frac{1}{q^2} +1 \right)}{(t-x)^2}  + b(x) \frac{\frac{1}{q^2}}{(t-x)} + c \right]}_{J(x,t)} dt = 0
\end{equation}
We need to show that
\begin{equation}
    I(x,t) J(x,t) = \frac{\partial H}{\partial t}
\end{equation}
We take
\begin{equation}
    H = I(x,t) M(x,t); \ \ \ \ \text{with} \ \ \ \ \ M(x,t) = \frac{ k \ t(t-1)}{(t-x)}; \ \ \ \text{where k is a cosnstant}
\end{equation}
Then,
\begin{equation}
    I(x,t) J(x,t) =  \frac{\partial H}{\partial t} = I(x,t) \left[ \frac{-1+\frac{3}{q^2}}{t} - \frac{\frac{1}{q^2}}{t-1} - \frac{\frac{1}{q^2}}{t-x}  \right] M(x,t) + I(x,t) \frac{\partial M}{\partial t}
    \label{corrrr}
\end{equation}
From equation (\ref{corrrr}), we obtain
\begin{equation}
     a(x) \frac{\frac{1}{q^2} \left( \frac{1}{q^2} +1 \right)}{(t-x)^2}  + b(x) \frac{\frac{1}{q^2}}{(t-x)} + c =  \left( \frac{-1+\frac{3}{q^2}}{t} - \frac{\frac{1}{q^2}}{t-1} - \frac{\frac{1}{q^2}}{t-x}  \right)M + \frac{\partial M}{\partial t}  \ \ \text{(for the taken M above)}
     \label{corr2}
\end{equation}
Now, to compare the coefficients of $t^2$, $t^1$, and $t^0$ on both sides, we multiply equation (\ref{corr2}) by $(t-x)^2$. After simplification, we obtain
\begin{equation}
    \begin{aligned}
        & t^2: \ \ \frac{k}{q^2} -c = 0,  \ \ \text{so}, c = \frac{k}{q^2}  \\ & t^1: \ \ b(x) = \frac{k-kx-\frac{2k}{q^2}}{\frac{1}{q^2}} \\& t^0:  \ \ a(x) = \frac{kx(1-x)}{\frac{1}{q^2}}
    \end{aligned}
\end{equation}
now taking $k=\frac{1}{q^2}$, we get
\begin{equation}
    a(x) = x(1-x); \ \ \ b(x) = 1-x-\frac{2}{q^2}; \ \ \ c = \frac{1}{q^4}
\end{equation}
Compared to the general form of the hypergeometric equation,
\begin{equation}
     A= \frac{1}{q^2}; \ \ \ B= -\frac{1}{q^2}; \ \ \ C = 1-\frac{2}{q^2}
\end{equation}
which allows us to write the final solution as (\ref{fil}).\\
\newline
Now, coming back to the integral representation, there is an important observation of expression (\ref{60}) can lead us to the idea of matrix models. We can rewrite (\ref{60}) as
\begin{equation}
     \oint dt \left[ t^{\alpha_1} \cdot (t-x)^{\alpha_2} \cdot (t-1)^{\alpha_3}    \right] \ \ \ \text{with} \ \alpha_1  = \frac{1-k}{k+2}; \ \alpha_2 = \alpha_3 =  - \frac{1}{k+2}
\end{equation}
Which can be further written as
\begin{equation}
    \oint dt \ e^{V_0 (t)} \ \ \ \ \text{with} \ \ \ \ V_0(t) = \alpha_1 \log t + \alpha_2 \log (t-x) + \alpha_3 \log(t-1)
    \label{62}
\end{equation}
As we have used only one screening charge here, we have one contour integral, and this particular interpretation can be thought of as a one matrix model with the given logarithmic potential (\ref{62}). This observation connects such an integral to Penner-type matrix models \cite{Penner}, which have many applications in the study of Toda and topological string theories \cite{DV}.

\subsection{A more general picture of four-point function}
In a general case, we are allowed to take different values of $j$ and $m$ to label the vertex operators, if they follow several conditions. Namely, we can calculate the correlator of the following type
\begin{equation}
    \Bigr \langle \tilde{V}_{j_1,m_1 } (0) \  V_{j_2,m_2} (1) \ V_{j_3,m_3} (x) \  V_{j_4,m_4 } (\infty) Q^L(t) V_s(R) \Bigr \rangle
    \label{855}
\end{equation}
where $L$ is the number of screening charges required to fulfill the charge neutrality condition. In our previous example we have seen the case $j_1 = j_2 = j_3 = j_4$, when all the $j$'s are fundamental, and only one screening charge was required $(L=1)$. As different combinations of $(j_i,m_i)$ at different operators also provide a non-zero correlator, they need to satisfy some conditions.\
\newline
The spin representation $j$ must follow the fusion rules for the algebra. They tell us how the primary fields fuse with each other in the OPE. If we have two primary fields $V_{j_1}$ and $V_{j_2}$, which are labeled by the representation $j$, then their fusion rules read as
\begin{equation}
    V_{j_1} \times V_{j_2} = \sum_{j_s} \mathcal{N}_{j_1 j_2}^{j_s} V_{j_s}
\end{equation}
where $\mathcal{N}_{j_1 j_2}^{j_s} \in \mathbb{Z}_{\geq 0} $ is the fusion coefficient. This is defined as
\begin{equation}
\mathcal{N}_{j_1 j_2}^{j_s} =
\begin{cases}
1 & \text{if } \left| j_1 - j_2 \right| \leq j_s \leq \min\left\{j_1 + j_2,\ 2k - j_1 - j_2\right\} \\
  & \text{and } j_1 + j_2 + j_s = \text{even integer} \\
0 & \text{otherwise}
\end{cases}
\label{877}
\end{equation}
The next step is to fuse $V_{j_s}$ with $V_{j_3}$ and must have to get $V_{j_4}$. From (\ref{877}), we see that $j_s$ might be several. If we get at least one $j_s$, such that $V_{j_s}$ fuse with $V_{j_3}$ gives $V_{j_4}$, then we get a non-zero conformal block. Here, this particular choice of fusing $j_1$ and $j_2$ first is called \textit{$s-$channel}. We could definitely try other possible choices as well. In that case, we have two more channels.
\begin{itemize}
    \item \textit{$t-$ channel:} fusing $j_2$, $j_3$ first to get $j_t$ and then fusing $j_t$, $j_1$ to get $j_{4}$
    \item \textit{$u-$ channel:} fusing $j_1$, $j_3$ first to get $j_u$ and then fusing $j_u$, $j_2$ to get $j_{4}$
\end{itemize}
Now, if in one channel the fusion rules forbid the fusion, we need to check the other channels as well. If fusion is restricted in all channels, then for that particular combination of $j$ the corresponding conformal block will vanish. In order to obtain a non-vanishing conformal block, it is therefore necessary that the $j$’s satisfy these fusion rules. Before looking to another example, let us summarize all the conditions that $(j_i,m_i)$ must have to satisfy in order to get a non-zero correlator.
\tcbset{colframe=gray, colback=white, boxrule=0.5pt, arc=2mm, boxsep=4pt}
\begin{tcolorbox}[title=Conditions on $j$ and $m$]
\begin{itemize}
    \item [1.] For a level $k$, choose the following representations $0 \leq \{j_i\}_{i=1}^4 \leq \frac{k}{2}$.
    \item[2.]  Check the fusion rules for $\{j_i\}_{i=1}^4$. If it doesn't satisfy, choose another combination of $\{j_i\}_{i=1}^4$. If it satisfies,
    \item[3.] Take $\{m_i\}_{i=1}^4 \in [-j,j] \cap \mathbb{Z}$ such that $\sum_{i=1}^4 m_i = 0$.
    \item[4.] Make four pairs of $(j,m)$. Note that, permuting $\{m_i\}_{i=1}^4$ between each other doesn't affect the final answer for the integral representation.
    \item[5.] To satisfy the charge neutrality condition, the number of screening operators $L$ needs to be $L \in Z_{\geq 0}$. As in the correlator (\ref{855})  we leveled the reflected operator $\tilde{V}$ by $j_1$, $L$ must be $L = j_2 + j_3 + j_4 - j_1$.
\end{itemize}
\end{tcolorbox}
After following the conditions above, one can proceed to construct the vertex operators for the corresponding representations and then derive the DF integral using the method described above. Let us look at several examples of such DF integrals for different cases.\\
\newline
For $k=10$, we choose the following pairs of $(j,m):$
\begin{equation}
   ( j_1,m_1 ) = \left(\frac{3}{2}, - \frac{3}{2} \right); \ \ ( j_2,m_2 ) = \left(\frac{3}{2}, \frac{3}{2} \right); \ \ ( j_3,m_3 ) = \left(\frac{5}{2}, - \frac{5}{2} \right);\ \ ( j_4,m_4 ) = \left(\frac{5}{2}, \frac{5}{2} \right);\ \
\end{equation}
Number of required screening operators: $L=5$
\begin{equation}
\begin{aligned}
    & \Bigr \langle \tilde{V}_{\left(\frac{3}{2}, - \frac{3}{2} \right)} (0) \  V_{\left(\frac{3}{2}, \frac{3}{2} \right)} (1) \ V_{\left(\frac{5}{2}, -\frac{5}{2} \right)} (x) \  V_{\left(\frac{5}{2}, \frac{5}{2} \right) } (\infty) Q^5(t) V_s(R) \Bigr \rangle =\\ &= \mathcal{G}(q,x) \oint \prod_{i=1}^5 dt_i \prod_{1 \leq i < j \leq 5} (t_i - t_j)^{- \frac{2}{q^{2}}} \prod_{i=1}^{5} \left[
     \ t_i^{-1 + \frac{5}{q^{2}}} \ (t_i - x)^{- \frac{5}{q^{2}}} \ (t_i - 1)^{- \frac{3}{q^{2}}}
    \right]
\end{aligned}
\end{equation}
By recalling that $q^2 = k+2$ and $k = 10$, we can rewrite this contour integral so that it depends only on $t$. Let us look at another example: for $k = 8$, we choose the following pairs of $(j,m)$:
\begin{equation}
   ( j_1,m_1 ) = (2,-2); \ \ ( j_2,m_2 ) = (4,-4); \ \ ( j_3,m_3) = (4,4); \ \ ( j_4,m_4 ) = (2,2); \ \
\end{equation}
Number of required screening operators: $L=8$
\begin{equation}
\begin{aligned}
    & \Bigr \langle \tilde{V}_{2,-2 } (0) \  V_{4,-4} (1) \ V_{4,4} (x) \  V_{2,2} (\infty) Q^8(t) V_s(R) \Bigr \rangle = \\ &= \mathcal{G}(q,x) \oint \prod_{i=1}^8 dt_i  \prod_{1 \leq i < j \leq 8} (t_i - t_j)^{- \frac{2}{q^{2}}} \prod_{i=1}^{8} \left[
    - \ t_i^{-1 + \frac{6}{q^{2}}} \ (t_i - x)^{- \frac{8}{q^{2}}} \ (t_i - 1)^{- \frac{8}{q^{2}}}
    \right]
\end{aligned}
\end{equation}
Satisfying the conditions above, a general formula for the DF integral representation of correlators in $\hat{sl}(2)_k$ WZW model:
\begin{equation}
\begin{aligned}
  & \Bigr \langle \tilde{V}_{j_1,m_1 } (0) \  V_{j_2,m_2} (1) \ V_{j_3,m_3} (x) \  V_{j_4,m_4 } (\infty) Q^L(t) V_s(R) \Bigr \rangle = \\ &= \boxed{ \mathcal{G} (q,x) \oint \prod_{i=1}^L dt_i \prod_{i<j}^L (t_i - t_j)^{-\frac{2}{q^2}} \prod_{i=1}^L t_i^{\frac{2(1+j_1)}{q^2}} (t_i-1)^{-\frac{2j_2}{q^2}} (t_i - x)^{-\frac{2j_3}{q^2}}}
\end{aligned}
\end{equation}
Therefore, we see that for a given level $k$, any combination of $j$ and $m$ satisfying the rules above yields a non-vanishing correlator. Changing $m$ does not affect the integral representation of the correlator as long as $\sum_i m_i = 0$ holds. Hence, the correlator should be invariant under the $SU(2)$ group action. Let us now consider this in more detail.

\subsection{Global and local action of $SU(2)$ }
For a fixed $j$ satisfying the fusion rules, each allowed choice of $(j,m)$ corresponds to a conformal block of the model. We can now act with $SU(2)$ on the vertex operator $V_{j,m}$, which simply changes the value of $m$ in the state $\ket{j,m}$. Roughly speaking, we rotate the states that label the vertex operators, and this generates an internal $SU(2)$ symmetry. For $g \in SU(2)$, the correlator is invariant under this global $SU(2)$ action. Let us demonstrate this in explicit examples.\\
As a simple example, let us consider again the spin-$\frac{1}{2}$ representation where we have two possible states $\ket{\uparrow} = \begin{pmatrix}
1  \\
0 \\
\end{pmatrix}$ and $\ket{\downarrow} = \begin{pmatrix}
0  \\
1 \\
\end{pmatrix}$. These states will corresponds to the vertex operators $V_{\left( \frac{1}{2},\frac{1}{2} \right)}$ and $V_{\left(\frac{1}{2}, -\frac{1}{2}\right)}$ respectively. We choose $g_{\pi} =e^{-i\pi J_y}$, which is the rotation by $\pi$ around $y$-axis. Note that we could consider other axis also but showing the rotation with respect to one angle is enough in this case.
\begin{equation}
    J_y = \frac{1}{2} \sigma_y = \frac{1}{2} \begin{pmatrix}
0 & -i \\
i & 0 \\
\end{pmatrix}; \ \ \text{then} \ \ g_{\pi} = e^{-i\pi J_y} = -i\sigma_y = \begin{pmatrix}
0 & -1 \\
1 & 0 \\
\end{pmatrix}
\end{equation}
Then we see
\begin{equation}
    g_{\pi}  \cdot \ket{\uparrow} = \begin{pmatrix}
0 & -1 \\
1 & 0 \\
\end{pmatrix} \begin{pmatrix}
1  \\
0  \\
\end{pmatrix} = \begin{pmatrix}
0  \\
1  \\
\end{pmatrix} = \ket{\downarrow}, \ \ \ \ \ \ \  g_{\pi}  \cdot \ket{\downarrow} = \begin{pmatrix}
0 & -1 \\
1 & 0 \\
\end{pmatrix} \begin{pmatrix}
0  \\
1  \\
\end{pmatrix} = \begin{pmatrix}
-1  \\
0  \\
\end{pmatrix} = - \ket{\uparrow}
\end{equation}
Which corresponds to
\begin{equation}
    g_{\pi}  \cdot V_{\left( \frac{1}{2},\frac{1}{2} \right)} (z_i) = V_{ \left( \frac{1}{2},-\frac{1}{2}\right)} (z_i), \ \ \ \ \ \ g_{\pi} \cdot V_{\left( \frac{1}{2},-\frac{1}{2} \right)} (z_i) = - V_{\left( \frac{1}{2},\frac{1}{2}\right)} (z_i)
\end{equation}
Now, by rotating all the states by the same angle $\pi$, we can calculate the correlate as
\begin{equation}
\begin{aligned}
     & \Bigr \langle \left( g_{\pi} \cdot \tilde{V}_{\left(\frac{1}{2},-\frac{1}{2}\right)} (0) \right) \left( g_{\pi}  \cdot V_{\left( \frac{1}{2},\frac{1}{2}\right)} (1) \right) \left( g_{\pi} \cdot V_{\left( \frac{1}{2},\frac{1}{2}\right)} (x) \right) \left( g_{\pi} \cdot V_{\left( \frac{1}{2},-\frac{1}{2}\right)}(\infty) \right) \  Q(t) \ V_s(R) \Bigr \rangle =\\    & \Bigr \langle \left(- \tilde{V}_{\left(\frac{1}{2},\frac{1}{2}\right)} (0) \right) \left( V_{\left( \frac{1}{2},-\frac{1}{2}\right)} (1) \right) \left( V_{\left( \frac{1}{2},-\frac{1}{2}\right)} (x) \right) \left( - V_{\left( \frac{1}{2},\frac{1}{2}\right)}(\infty) \right) \  Q(t) \ V_s(R) \Bigr \rangle =  \\&  =  \Bigr \langle  \tilde{V}_{\left(\frac{1}{2},-\frac{1}{2}\right)} (0) \ V_{\left( \frac{1}{2},\frac{1}{2}\right)} (1) \ V_{\left( \frac{1}{2},\frac{1}{2}\right)} (x) \ V_{\left( \frac{1}{2},-\frac{1}{2}\right)}(\infty) \  Q(t) \ V_s(R) \Bigr \rangle
    \end{aligned}
\end{equation}
We see that correlator is invariant under $\pi$-rotation of each states in the vertex operators. Now we consider several possible cases where we might look for the $SU(2)$ symmetry. We are allowed to take different $j$-spin representations at different points. We can also rotate all the states by a different angle $\beta$. To gain an overview of a more general picture, we utilize the concept of the Wigner $d$-matrix. The action of a group element $g_{\beta}$ on the vertex operator
\begin{equation}
    g_{\beta} \cdot V_{j,m} = \sum_{m' = - j}^j d_{m',m}^{(j)} (\beta) V_{j,m'}
\end{equation}
where $d_{m',m}^{(j)} (\beta)$ is the Wigner small $d$-matrix for a rotation by $\beta$ for any spin $j$. This matrix provides the coefficient of each state under rotation. To calculate that, we use the matrix exponential method. Let us consider the rotation under the $y$-axis. Then the angular momentum operator $J_y$, a $(2j+1) \times (2j+1)$ matrix, depending on the raising and lowering operator, can be written as
\begin{equation}
    \braket{j,m' | J_y | j,m} = \frac{1}{2\pi} \left[ \sqrt{(j-m) (j+m+1)}\delta_{m',m+1} - \sqrt{(j+m) (j-m+1)}\delta_{m',m-1} \right]
\end{equation}
Then $d_{m',m}^{(j)} (\beta)$ defined as
\begin{equation}
    d_{m'm}^{(j)} (\beta) =  e^{-i \beta J_y}
\end{equation}
which can be easily obtained by expanding matrix exponential. To show invariance under the group action $SU(2)$, we use the Ward identities for infinitesimal rotations $\beta$. For this, we expand  $d_{m',m}^{(j)} (\beta)$ in powers of $\beta$. Then the Taylor expansion around the $\beta=0$ gives
\begin{equation}
    g_{\beta} \cdot V_{j,m}= \left.\sum_{m'} \left( \delta_{m,m'} + \beta \frac{d}{d\beta} \ d_{m',m}^{(j)} (\beta) \right|_{\beta=0} + \mathcal{O}(\beta^2) \right) V_{j,m'}
\end{equation}
Then applying the infinitesimal rotation to each operator at our dedicated four points:
\begin{equation}
    \begin{aligned}
& g_{\beta} \cdot \tilde{V}(0)_{j_1,m_1} = \tilde{V}_{j_1,m_1} (0) + \left. \beta \left( \frac{d}{d\beta} d_{m_1',m_1}^{j_1} \right|_{\beta=0} +\mathcal{O}(\beta^2)\right) \tilde{V}_{j_1,m_1'} (0) \\& g_{\beta} \cdot V(1)_{j_2,m_2} = V_{j_2,m_2} (1) + \left. \beta \left( \frac{d}{d\beta} d_{m_2',m_2}^{j_2} \right|_{\beta=0} + \mathcal{O}(\beta^2)\right) V_{j_2,m_2'} (1) \\& g_{\beta} \cdot V(x)_{j_3,m_3} = V_{j_3,m_3} (x) + \left. \beta \left( \frac{d}{d\beta} d_{m_3',m_3}^{j_3} \right|_{\beta=0} + \mathcal{O}(\beta^2)\right) V_{j_3,m_3'} (x) \\& g_{\beta} \cdot V(\infty)_{j_4,m_4} = V_{j_4,m_4} (\infty) + \left. \beta \left( \frac{d}{d\beta} d_{m_4',m_4}^{j_4} \right|_{\beta=0} + \mathcal{O}(\beta^2) \right) V_{j_4,m_4'} (\infty)
    \end{aligned}
\end{equation}
If the theory has a $SU(2)$ global invariance then
\begin{equation}
    \begin{aligned}
   & \Bigr \langle \tilde{V}_{j_1,m_1' } (0) \  V_{j_2,m_2} (1) \ V_{j_3,m_3} (x) \  V_{j_4,m_4 } (\infty) Q^L(t) V_s(R) \Bigr \rangle \left( \left. \beta \left( \frac{d}{d\beta} d_{m_1',m_1}^{j_1} \right|_{\beta=0} +\mathcal{O}(\beta^2) \right)\right)   + \\&+   \Bigr \langle \tilde{V}_{j_1,m_1 } (0) \  V_{j_2,m_2'} (1) \ V_{j_3,m_3} (x) \  V_{j_4,m_4 } (\infty) Q^L(t) V_s(R) \Bigr \rangle \left( \left. \beta \left( \frac{d}{d\beta} d_{m_2',m_2}^{j_2} \right|_{\beta=0} +\mathcal{O}(\beta^2) \right)\right)  + \\&+   \Bigr \langle \tilde{V}_{j_1,m_1 } (0) \  V_{j_2,m_2} (1) \ V_{j_3,m_3'} (x) \  V_{j_4,m_4 } (\infty) Q^L(t) V_s(R) \Bigr \rangle \left( \left. \beta \left( \frac{d}{d\beta} d_{m_3',m_3}^{j_2} \right|_{\beta=0} +\mathcal{O}(\beta^2) \right)\right)   + \\&+   \Bigr \langle \tilde{V}_{j_1,m_1 } (0) \  V_{j_2,m_2} (1) \ V_{j_3,m_3} (x) \  V_{j_4,m_4' } (\infty) Q^L(t) V_s(R) \Bigr \rangle \left( \left. \beta \left( \frac{d}{d\beta} d_{m_4',m_4}^{j_4} \right|_{\beta=0} +\mathcal{O}(\beta^2) \right)\right)  = \\& = \  \Bigr \langle g_{\beta} \cdot \tilde{V}_{j_1,m_1 } (0) \ g_{\beta} \cdot  V_{j_2,m_2} (1) \ g_{\beta} \cdot V_{j_3,m_3'} (x) \  g_{\beta} \cdot  V_{j_4,m_4 } (\infty) Q^L(t) V_s(R) \Bigr \rangle \  -  \\&  - \ \Bigr \langle \tilde{V}_{j_1,m_1 } (0) \  V_{j_2,m_2} (1) \ V_{j_3,m_3} (x) \  V_{j_4,m_4' } (\infty) Q^L(t) V_s(R) \Bigr \rangle = 0
    \end{aligned}
    \label{10000}
\end{equation}
To be globally invariant from the above expression, it must hold
\begin{equation}
    \beta ( \mathfrak{D}) = 0
\end{equation}
and for any $\beta$, the big expression $\mathfrak{D}$ must have to be zero. Let's illustrate several example as possible cases.\\
\newline
\boxed{ \textbf{Case 1: $\Bigr \langle g_{\beta} \cdot \tilde{V}_{ \left( \frac{1}{2},\frac{1}{2} \right)} (0) \ g_{\beta} \cdot  V_{ \left(\frac{1}{2}, -\frac{1}{2} \right)} (1) \ g_{\beta} \cdot V_{\left(\frac{1}{2}, -\frac{1}{2}\right)} (x) \  g_{\beta} \cdot  V_{\left(\frac{1}{2},\frac{1}{2} \right) } (\infty) Q(t) V_s(R) \Bigr \rangle$}}\\
\newline
The Wigner small-d matrix \( d^{(1/2)}(\beta) \) is:

\begin{equation}
d^{(1/2)}(\beta) =
\begin{pmatrix}
\cos\left( \frac{\beta}{2} \right) & -\sin\left( \frac{\beta}{2} \right) \\
\sin\left( \frac{\beta}{2} \right) & \cos\left( \frac{\beta}{2} \right)
\end{pmatrix}
\end{equation}

Derivative of \( d^{(1/2)}(\beta) \) at \( \beta = 0 \):
\[
\frac{d}{d\beta} d^{(1/2)}(\beta) \Big|_{\beta=0} =
\begin{pmatrix}
0 & -\frac{1}{2} \\
\frac{1}{2} & 0
\end{pmatrix}
\]
Then the action of the group element rotate the states as
\begin{equation}
\begin{aligned}
    &g_{\beta} \cdot V_{ \left( \frac{1}{2},\frac{1}{2} \right)} (z_i) = V_{ \left( \frac{1}{2},\frac{1}{2} \right)} (z_i) + \beta \cdot \left( \frac{1}{2} V_{ \left( \frac{1}{2},-\frac{1}{2} \right)} (z_i)\right) + \mathcal{O}(\beta^2), \\&  g_{\beta} \cdot V_{ \left( \frac{1}{2},-\frac{1}{2} \right)} (z_i) = V_{ \left( \frac{1}{2},-\frac{1}{2} \right)} (z_i) - \beta \cdot \left( \frac{1}{2} V_{ \left( \frac{1}{2},\frac{1}{2} \right)} (z_i)\right)  + \mathcal{O}(\beta^2)
    \end{aligned}
\end{equation}
Putting these rotated states into the correlator and simplifying them gives
\begin{equation}
    \beta \left( \frac{1}{2} \Bigr \langle  \tilde{V}_{\left(\frac{1}{2},\frac{1}{2}\right)} (0) \ V_{\left( \frac{1}{2},-\frac{1}{2}\right)} (1) \ V_{\left( \frac{1}{2},-\frac{1}{2}\right)} (x) \ V_{\left( \frac{1}{2},-\frac{1}{2}\right)}(\infty) \  Q(t) \ V_s(R) \Bigr \rangle \right) = C_{\text{rotated}} - C_{\text{original}} = 0
    \label{1040}
\end{equation}
We see that (\ref{1040}) holds for any $\beta$ because the correlator here vanishes due to not satisfying $\sum_i m_i = 0$.\\
\newline
\textbf{So, the global $SU(2)$ symmetry preserves.}\\
\newline
\boxed{ \textbf{Case 2: $\Bigr \langle g_{\beta} \cdot \tilde{V}_{ \left( \frac{1}{2},-\frac{1}{2} \right)} (0) \ g_{\beta} \cdot  V_{ \left( 1,-1\right)} (1) \ g_{\beta} \cdot V_{\left(1,1 \right)} (x) \  g_{\beta} \cdot  V_{\left(\frac{1}{2},\frac{1}{2} \right) } (\infty) Q^2(t) V_s(R) \Bigr \rangle$}}\\
\newline
The Wigner small-d matrix \( d^{(1)}(\beta) \) is:
\[
d^{(1)}(\beta) =
\begin{pmatrix}
\cos^2\left( \frac{\beta}{2} \right) & -\frac{\sin \beta}{\sqrt{2}} & \sin^2\left( \frac{\beta}{2} \right) \\
\frac{\sin \beta}{\sqrt{2}} & \cos \beta & -\frac{\sin \beta}{\sqrt{2}} \\
\sin^2\left( \frac{\beta}{2} \right) & \frac{\sin \beta}{\sqrt{2}} & \cos^2\left( \frac{\beta}{2} \right)
\end{pmatrix}
\]
Derivative of \( d^{(1)}(\beta) \) at \( \beta = 0 \):
\[
\left. \frac{d}{d\beta} d^{(1)}(\beta) \right|_{\beta=0} =
\begin{pmatrix}
0 & -\frac{1}{\sqrt{2}} & 0 \\
\frac{1}{\sqrt{2}} & 0 & -\frac{1}{\sqrt{2}} \\
0 & \frac{1}{\sqrt{2}} & 0
\end{pmatrix}
\]
Then the action of the group element rotate the states as
\begin{equation}
\begin{aligned}
    &g_{\beta} \cdot V_{ \left( 1,1 \right)} (z_i) = V_{ \left( 1,1\right)} (z_i) - \beta \cdot \left( \frac{1}{\sqrt{2}} V_{ \left( 1,0 \right)} (z_i)\right) + \mathcal{O}(\beta^2), \\&  g_{\beta} \cdot V_{ \left( 1,-1 \right)} (z_i) = V_{ \left( 1,-1\right)} (z_i) + \beta \cdot \left( \frac{1}{\sqrt{2}} V_{ \left( 1,0 \right)} (z_i)\right) + \mathcal{O}(\beta^2)
    \end{aligned}
\end{equation}
Again putting these rotated states into the correlator and simplifying them gives
\begin{equation}
    \beta \left( \frac{1}{2} \Bigr \langle  \tilde{V}_{\left(\frac{1}{2},-\frac{1}{2}\right)} (0) \ V_{\left( 1,-1\right)} (1) \ V_{\left( 1,1\right)} (x) \ V_{\left( \frac{1}{2},-\frac{1}{2}\right)}(\infty) \  Q^2(t) \ V_s(R) \Bigr \rangle \right) = C_{\text{rotated}} - C_{\text{original}} = 0
\end{equation}
\textbf{Again, the global $SU(2)$ symmetry preserves}.\\
\newline
\boxed{ \textbf{Case 3: $\Bigr \langle \left( g_{\beta} \cdot \tilde{V}_{ \left( \frac{1}{2},-\frac{1}{2} \right)} (0) \right)\  V_{ \left( 1,-1\right)} (1) \ V_{\left(1,1 \right)} (x) \ V_{\left(\frac{1}{2},\frac{1}{2} \right) } (\infty) Q^2(t) V_s(R) \Bigr \rangle$}}\\
\newline
We rotated only one state at point $0$. In this case we get
\begin{equation}
    -\beta \left( \frac{1}{2} \Bigr \langle  \tilde{V}_{\left(\frac{1}{2},\frac{1}{2}\right)} (0) \ V_{\left( 1,-1\right)} (1) \ V_{\left( 1,1\right)} (x) \ V_{\left( \frac{1}{2},\frac{1}{2}\right)}(\infty) \  Q^2(t) \ V_s(R) \Bigr \rangle \right) = C_{\text{rotated}} - C_{\text{original}} = 0
\end{equation}
\textbf{Global $SU(2)$ symmetry preserves.}\\
\newline
\boxed{ \textbf{Case 4: $\Bigr \langle \left( g_{\beta} \cdot \tilde{V}_{ \left( \frac{1}{2},-\frac{1}{2} \right)} (0) \right)\  V_{ \left( 1,-1\right)} (1) \  \left( g_{\beta} \cdot V_{\left(1,1 \right)} (x) \right) \ V_{\left(\frac{1}{2},\frac{1}{2} \right) } (\infty) Q^2(t) V_s(R) \Bigr \rangle$}}\\
\newline
We rotated two states at points $0$ and $x$ by the same angle $\beta$. In this case we get
\begin{equation}
    -\beta \left( \frac{1}{\sqrt{2}} \Bigr \langle  \tilde{V}_{\left(\frac{1}{2},-\frac{1}{2}\right)} (0) \ V_{\left( 1,-1\right)} (1) \ V_{\left( 1,0\right)} (x) \ V_{\left( \frac{1}{2},\frac{1}{2}\right)}(\infty) \  Q^2(t) \ V_s(R) \Bigr \rangle \right) = C_{\text{rotated}} - C_{\text{original}} = 0
\end{equation}
\textbf{Global $SU(2)$ symmetry preserves.}\\
\newline
\boxed{ \textbf{Case 5: $\Bigr \langle g_{\beta_1} \cdot \tilde{V}_{ \left( \frac{1}{2},-\frac{1}{2} \right)} (0) \ g_{\beta_2} \cdot  V_{ \left( 1,-1\right)} (1) \ g_{\beta_3} \cdot V_{\left(1,1 \right)} (x) \  g_{\beta_4} \cdot  V_{\left(\frac{1}{2},\frac{1}{2} \right) } (\infty) Q^2(t) V_s(R) \Bigr \rangle$}}\\
\newline
In this case, we considered the local $SU(2)$ symmetry. We rotated each state at each point by different angles $\beta_i$. In this case, to preserve the symmetry, we must have
\begin{equation}
\begin{aligned}
    &-\frac{\beta_2 \beta_3}{2} \left( \Bigr \langle  \tilde{V}_{\left(\frac{1}{2},-\frac{1}{2}\right)} (0) \ V_{\left( 1,0\right)} (1) \ V_{\left( 1,0\right)} (x) \ V_{\left( \frac{1}{2},\frac{1}{2}\right)}(\infty) \  Q^2(t) \ V_s(R) \Bigr \rangle \right) + \\& + \frac{\beta_2 \beta_4}{2 \sqrt{2}} \left( \Bigr \langle  \tilde{V}_{\left(\frac{1}{2},-\frac{1}{2}\right)} (0) \ V_{\left( 1,0\right)} (1) \ V_{\left( 1,1\right)} (x) \ V_{\left( \frac{1}{2},-\frac{1}{2}\right)}(\infty) \  Q^2(t) \ V_s(R) \Bigr \rangle \right) + \\& + \frac{\beta_1 \beta_3}{2 \sqrt{2}} \left( \Bigr \langle  \tilde{V}_{\left(\frac{1}{2},\frac{1}{2}\right)} (0) \ V_{\left( 1,-1\right)} (1) \ V_{\left( 1,0\right)} (x) \ V_{\left( \frac{1}{2},\frac{1}{2}\right)}(\infty) \  Q^2(t) \ V_s(R) \Bigr \rangle \right) - \\& - \frac{\beta_1 \beta_4}{4} \left( \Bigr \langle  \tilde{V}_{\left(\frac{1}{2},\frac{1}{2}\right)} (0) \ V_{\left( 1,-1\right)} (1) \ V_{\left( 1,1\right)} (x) \ V_{\left( \frac{1}{2},-\frac{1}{2}\right)}(\infty) \  Q^2(t) \ V_s(R) \Bigr \rangle \right)  + \\& + \frac{ \beta_1 \beta_2 \beta_3 \beta_4}{8} \left( \Bigr \langle  \tilde{V}_{\left(\frac{1}{2},\frac{1}{2}\right)} (0) \ V_{\left( 1,0\right)} (1) \ V_{\left( 1,0\right)} (x) \ V_{\left( \frac{1}{2}, - \frac{1}{2}\right)}(\infty) \  Q^2(t) \ V_s(R) \Bigr \rangle \right)  = \\& =  C_{\text{rotated}} - C_{\text{original}} = 0
    \end{aligned}
    \label{1090}
\end{equation}
As we have seen before, the value of the integral expression of the non-vanishing correlator depends directly on $j$ and not on $m$. So for all these 5 correlators in (\ref{1090}), we get the same value. We can further simplify this as
\begin{equation}
   \left( -\frac{\beta_2 \beta_3}{2} +\frac{\beta_2 \beta_4}{2 \sqrt{2}} + \frac{\beta_1 \beta_3}{2 \sqrt{2}} - \frac{\beta_1 \beta_4}{4 } + \frac{\beta_1 \beta_2 \beta_3 \beta_4}{8 }  \right) \left( \Bigr \langle  \tilde{V}_{\left(\frac{1}{2},-\frac{1}{2}\right)} (0) \ V_{\left( 1,0\right)} (1) \ V_{\left( 1,0\right)} (x) \ V_{\left( \frac{1}{2},\frac{1}{2}\right)}(\infty) \  Q^2(t) \ V_s(R) \Bigr \rangle \right) = 0
\end{equation}
The correlator here is non-zero. So, to hold this equation, the combination of $\beta_i$ in the coefficient must have to be zero for any arbitrary $\beta$. But this does not happen, and it is a sign of the absence of local $SU(2)$ invariance. So we conclude that \\
\newline
\textbf{Local $SU(2)$ symmetry breaks.}\\
\newline
Now, this symmetry imposes a condition on the selection rules of $j$, which relates the fusion rules we have seen above. To get a non-zero correlator and observe the $SU(2)$ symmetry, the coupled state of all four spins $j$ should belong to the trivial representation of $sl(2)$. To check this for different-$j$ at different points, we decompose the tensor product of these spin-$j$ into irreducible representations. This decomposition must have to contain at least one singlet, which is a one dimensional subspace in the trivial representation of $sl(2)$. Otherwise, the global symmetry will break. To proceed, the tensor product of two such representations
\begin{equation}
    j_1 \otimes j_2 = \bigoplus_{j = |j_1 - j_2|}^{\,j_1 + j_2} j
\end{equation}
We can look at some examples to check. Lets take $j_1 = 1/2, j_2 = 1 , j_3 = 1, j_4 = 1/2$, then

\begin{equation}
    \begin{aligned}
       &  \frac{1}{2} \otimes 1 = \frac{1}{2} \oplus \frac{3}{2}, \ \ 1 \otimes \frac{1}{2} =  \frac{1}{2} \oplus \frac{3}{2} \\& \left( \frac{1}{2} \oplus \frac{3}{2} \right) \otimes  \left( \frac{1}{2} \oplus \frac{3}{2} \right) = \frac{1}{2} \otimes \frac{1}{2} \oplus \frac{1}{2} \otimes \frac{3}{2} \oplus \frac{3}{2} \otimes \frac{1}{2} \oplus\frac{3}{2} \otimes \frac{3}{2}= \\& = 0 \ \oplus 0 \ \oplus 1 \ \oplus 1  \ \oplus 1  \ \oplus 1  \ \oplus 2 \ \oplus 2 \ \oplus 3
    \end{aligned}
    \label{triv}
\end{equation}
The irreducible decomposition contains two singlets. They correspond to the trivial representation of $sl(2)$.  Now we act the generator of the algebra on that trivial representation using the co-product, which is defined as
\begin{equation}
    \rho (v \otimes w) = \rho_1 (v) \otimes w + v \otimes \rho_2(w) \ \ \ \text{where} \ \ \rho = \rho_1 \otimes \mathbf{1} + \mathbf{1} \otimes \rho_2
    \label{cop}
\end{equation}
Then we see that the two trivial representations in (\ref{triv}) come from the tensor product $\Big| \frac{1}{2} \Bigr \rangle  \otimes \Big| \frac{1}{2} \Bigr \rangle $ and $\Big| \frac{3}{2} \Bigr \rangle \otimes \Big| \frac{3}{2} \Bigr \rangle $. Now we need to create that singlet state using the generator of the algebra. Usually, all the elements of the algebra acting on the trivial representation give 0. For simplicity, now we can visualize the following example using the co-product (\ref{cop}) for the singlet state corresponding to $\frac{1}{2}$. We apply the Cartan generator to all the possible state settings.
\begin{equation}
    \begin{aligned}
        & J^3 \left( \bigg| \frac{1}{2} \biggr \rangle \otimes \bigg| \frac{1}{2} \biggr \rangle \right) = \bigg| \frac{1}{2} \biggr \rangle \otimes \bigg| \frac{1}{2} \biggr \rangle; \ \ \ \  J^3 \left( \bigg| - \frac{1}{2} \biggr \rangle \otimes \bigg| -\frac{1}{2} \biggr \rangle \right) = - \left( \bigg| \frac{1}{2} \biggr \rangle \otimes \bigg| \frac{1}{2} \biggr \rangle \right); \\& J^3 \left( \bigg| -\frac{1}{2} \biggr \rangle \otimes \bigg| \frac{1}{2} \biggr \rangle \right) = 0; \ \ \ \ J^3 \left( \bigg| \frac{1}{2} \biggr \rangle \otimes \bigg| -\frac{1}{2} \biggr \rangle \right) = 0 ;
    \end{aligned}
    \label{1150}
\end{equation}
We see two cases where the action of Cartan generator gives 0. So, they contribute to the singlet state. Now we make a linear combination of them as follows.
\begin{equation}
    A \left( \bigg| \frac{1}{2} \biggr \rangle \otimes \bigg| -\frac{1}{2} \biggr \rangle \right) + B \left( \bigg| \frac{1}{2} \biggr \rangle \otimes \bigg| -\frac{1}{2} \biggr \rangle  \right)
\end{equation}
Now, the raising operator acting on this linear combination gives 0. One can easily find $A = - B$. So finally, the singlet state is
\begin{equation}
    S= \frac{1}{\sqrt{2}} \left( \bigg| \frac{1}{2} \biggr \rangle \otimes \bigg| -\frac{1}{2} \biggr \rangle - \bigg|-\frac{1}{2} \biggr \rangle \otimes \bigg| \frac{1}{2} \biggr \rangle\right)
    \label{1170}
\end{equation}
Now its easy to check that the generators of $sl(2)$ acting on this gives 0.
\begin{equation}
    J^+\ket{S} = \frac{1}{\sqrt{2}} \left( \bigg|  \frac{1}{2}  \biggr \rangle \otimes \bigg|  \frac{1}{2}  \biggr \rangle  - \bigg| \frac{1}{2}  \biggr \rangle \otimes \bigg|  \frac{1}{2}  \biggr \rangle  \right) = 0
\end{equation}
\begin{equation}
    J^-\ket{S} = \frac{1}{\sqrt{2}} \left(\bigg|  -\frac{1}{2} \biggr \rangle\otimes \bigg|- \frac{1}{2} \biggr \rangle - \bigg|-\frac{1}{2}  \biggr \rangle \otimes \bigg|-\frac{1}{2} \biggr \rangle  \right) = 0
\end{equation}
\begin{equation}
\begin{aligned}
    J^3\ket{S} =&  \frac{1}{\sqrt{2}} \left[ \frac{1}{2} \left(\bigg| \frac{1}{2}\biggr \rangle \otimes \bigg|-\frac{1}{2} \biggr \rangle \right) - \frac{1}{2} \left(\bigg| \frac{1}{2} \biggr \rangle \otimes \bigg|- \frac{1}{2} \biggr \rangle \right) - \frac{1}{2} \left( \bigg|-\frac{1}{2} \biggr \rangle \otimes \bigg| \frac{1}{2} \biggr \rangle \right) + \frac{1}{2} \left(\bigg|-\frac{1}{2} \biggr \rangle \otimes \bigg| \frac{1}{2} \biggr \rangle\right) \right] = 0
    \end{aligned}
\end{equation}
Now let's consider a concrete example of different $j$. Let's take $j_1 = 4, j_2 = 3, j_3 = \frac{3}{2}, j_4 = \frac{1}{2}$. Here, the irreducible decomposition gives
\begin{equation}
    4 \otimes 3 \otimes\frac{3}{2} \otimes\frac{1}{2} = 0^2 \oplus 1^5 \oplus 2^7 \oplus 3^8  \oplus 4^8  \oplus 5^8 \oplus6^7 \oplus7^5 \oplus 8^3 \oplus 9
\end{equation}
So, the decomposition contains two singlet. Using the previous setup ( \ref{1150})-(\ref{1170}), one can create the singlet and check the action of the generators. Now containing singlets for these $j$ doesn't always necessarily give a non-zero correlator. They must have to obey the fusion rules for $\hat{sl}(2)_k$ (\ref{877}). By means of this, it depends on the level $k$ of the Kac-Moody algebra. For example, for this particular selected $j$, we can proceed to calculate the correlator. But the correlator will be non-zero only if the level of the level $k\geq8$. Then the vertex operators are
\begin{align}
    \tilde{V}_{(4;1)} (0)&= \exp\left( i \frac{5\sqrt{2}}{q} \, \phi - 6u-6iv) \right), &
     V_{(3; -1)} (1) &= \exp\left( i\frac{3\sqrt{2}}{q} \, \phi + 4u - 4iv \right)
     \label{kjh}
\end{align}
\begin{align}
    V_{\left( \frac{3}{2}, \frac{1}{2}  \right)} (x) &= \exp\left( i \frac{3}{\sqrt{2}q} \, \phi + u - iv\right), &
     V_{\left( \frac{1}{2}, -\frac{1}{2}  \right) } (\infty) &= \exp\left( i\frac{1}{\sqrt{2}q} \, \phi + u - iv \right)
     \label{kjh}
\end{align}
The number of screening charges here is $L = 3+\frac{3}{2}+\frac{1}{2}-4 = 1$. Then the correlator be
\begin{equation}
\begin{aligned}
  &\Bigr \langle \tilde{V}_{(4,1) } (0) \  V_{(3,-1)} (1) \ V_{\left( \frac{3}{2}, \frac{1}{2} \right) } (x) \  V_{\left( \frac{1}{2}, -\frac{1}{2} \right) } (\infty) Q(t) V_s(R) \Bigr \rangle =  \mathcal{G}(q,x) \oint dt \left[
    6 \ t^{-1 + \frac{10}{q^{2}}} \ (t - x)^{- \frac{3}{q^{2}}} \ (t - 1)^{- \frac{6}{q^{2}}}
    \right] \\&= \mathcal{G}(q,x) \oint dt \left[
    6 \ t^{-1 + \frac{10}{k+2 }} \ (t - x)^{- \frac{3}{k+2}} \ (t - 1)^{- \frac{6}{k+2}}
    \right]
\end{aligned}
\end{equation}
This integral expression is non-zero only when $k\geq 8$ due to the fusion rules (\ref{877}).\\
\newline
So finally, in the description above, we see that the correlator vanishes for the following reason. For selected spins (highest weights), the coupled state must be in the trivial representation, which ensures that $\sum_i m_i = 0$, and must obey the fusion rules of the WZW algebra. Then rotating the states changes only $m$, and after multiplication we obtain a correlator with a different choice of $m \in [-j,j]$. To proceed further, we first check whether $\sum_i m_i = 0$ is valid or not. If not, we remove this configuration, as it gives a vanishing correlator. If yes, then we keep it.

\section{Free field realization of $\hat{\mathfrak{sl}}(3)_k$ WZW model }
\subsection{Bosonization of $\hat{\mathfrak{sl}}(3)_k$ WZW model }
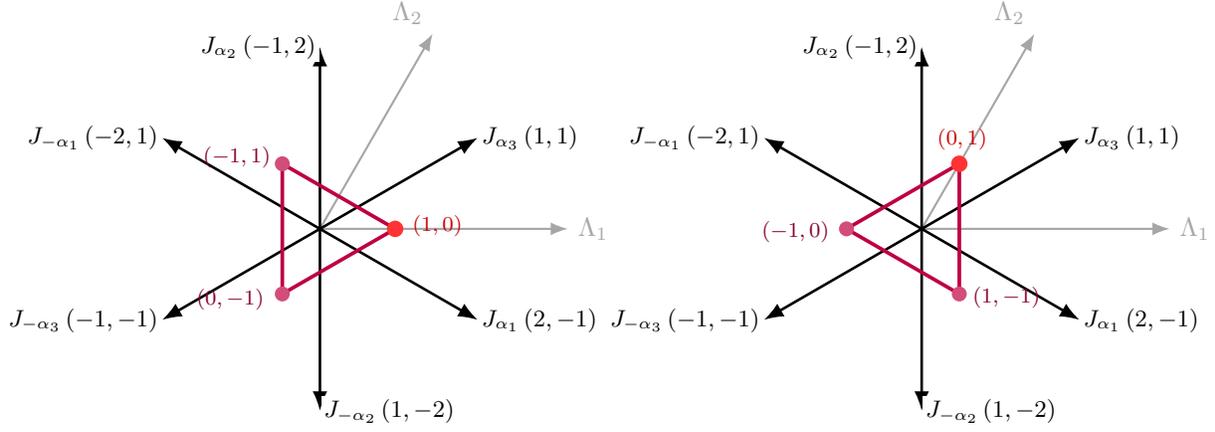
\begin{figure}[h!]
\centering
\begin{tikzpicture}[
  x={(1cm,0cm)}, y={(0.5cm,0.8660254cm)},
  rootarrow/.style = {->, line width=1.0pt},
  eroot/.style = {rootarrow, black},
  froot/.style = {rootarrow, black},
  weightpt/.style = {circle, fill=purple!70, inner sep=2pt},
  hwpt/.style = {circle, fill=red!80, inner sep=2.2pt},
  rlab/.style = {font=\small\itshape, fill=white, inner sep=1pt}
]

\begin{scope}[shift={(0,0)}]

\draw[->, thick, gray!70] (0,0) -- (3.3,0) node[anchor=west] {$\Lambda_1$};
\draw[->, thick, gray!70] (0,0) -- (0,3.0) node[anchor=south east] {$\Lambda_2$};

\def\R{1.4}
\draw[eroot] (0,0) -- ($(0,0)!\R!( 2,-1)$) node[rlab, anchor=west]  {$J_{\alpha_1}\,(2,-1)$};
\draw[eroot] (0,0) -- ($(0,0)!\R!(-1, 2)$) node[rlab, anchor=east] {$J_{\alpha_2}\,(-1,2)$};
\draw[eroot] (0,0) -- ($(0,0)!\R!( 1, 1)$) node[rlab, anchor=west]  {$J_{\alpha_3}\,(1,1)$};
\draw[froot] (0,0) -- ($(0,0)!\R!(-2, 1)$) node[rlab, anchor=east] {$J_{-\alpha_1}\,(-2,1)$};
\draw[froot] (0,0) -- ($(0,0)!\R!( 1,-2)$) node[rlab, anchor=west]  {$J_{-\alpha_2}\,(1,-2)$};
\draw[froot] (0,0) -- ($(0,0)!\R!(-1,-1)$) node[rlab, anchor=east] {$J_{-\alpha_3}\,(-1,-1)$};

\coordinate (A) at ( 1, 0);
\coordinate (B) at (-1, 1);
\coordinate (C) at ( 0,-1);
\draw[line width=1.4pt, purple] (A) -- (B) -- (C) -- cycle;
\draw[thick, purple, dashed] (A) -- (C);
\node[hwpt] at (A) {};
\node[weightpt] at (B) {};
\node[weightpt] at (C) {};
\node[font=\footnotesize, text=red!80!black, anchor=west]  at ($(A)+(0.08,0.05)$) {$(1,0)$};
\node[font=\footnotesize, text=purple!70!black, anchor=east] at ($(B)+(-0.08,0.10)$) {$(-1,1)$};
\node[font=\footnotesize, text=purple!70!black, anchor=east] at ($(C)+(-0.08,-0.08)$) {$(0,-1)$};
\end{scope}

\begin{scope}[shift={(8cm,0)}]
\draw[->, thick, gray!70] (0,0) -- (3.3,0) node[anchor=west] {$\Lambda_1$};
\draw[->, thick, gray!70] (0,0) -- (0,3.0) node[anchor=south east] {$\Lambda_2$};

\def\R{1.4}
\draw[eroot] (0,0) -- ($(0,0)!\R!( 2,-1)$) node[rlab, anchor=west]  {$J_{\alpha_1}\,(2,-1)$};
\draw[eroot] (0,0) -- ($(0,0)!\R!(-1, 2)$) node[rlab, anchor=east] {$J_{\alpha_2}\,(-1,2)$};
\draw[eroot] (0,0) -- ($(0,0)!\R!( 1, 1)$) node[rlab, anchor=west]  {$J_{\alpha_3}\,(1,1)$};
\draw[froot] (0,0) -- ($(0,0)!\R!(-2, 1)$) node[rlab, anchor=east] {$J_{-\alpha_1}\,(-2,1)$};
\draw[froot] (0,0) -- ($(0,0)!\R!( 1,-2)$) node[rlab, anchor=west]  {$J_{-\alpha_2}\,(1,-2)$};
\draw[froot] (0,0) -- ($(0,0)!\R!(-1,-1)$) node[rlab, anchor=east] {$J_{-\alpha_3}\,(-1,-1)$};

\coordinate (A) at ( 0, 1);
\coordinate (B) at ( 1,-1);
\coordinate (C) at (-1, 0);
\draw[line width=1.4pt, purple] (B) -- (A) -- (C) -- cycle;
\draw[thick, purple, dashed] (A) -- (C);
\node[hwpt] at (A) {};
\node[weightpt] at (B) {};
\node[weightpt] at (C) {};
\node[font=\footnotesize, text=red!80!black, anchor=south] at ($(A)+(0,0.08)$) {$(0,1)$};
\node[font=\footnotesize, text=purple!70!black, anchor=west]  at ($(B)+(0.08,-0.05)$) {$(1,-1)$};
\node[font=\footnotesize, text=purple!70!black, anchor=east]  at ($(C)+(-0.08,-0.05)$) {$(-1,0)$};

\end{scope}

\end{tikzpicture}
\caption{Weight diagrams of $\mathfrak{sl}_3$ for the fundamental $(1,0)$ and anti-fundamental $(0,1)$ representations.}
\label{fig:sl3-both}
\end{figure}

In case of $\hat{\mathfrak{sl}}(3)_k$, we follow the same procedure of $\hat{\mathfrak{sl}}(2)_k$. At first we construct the vertex operators from the bosonization and then we calculate the correlator. In case of $\hat{\mathfrak{sl}}(3)_k$, the bosonic fields $\chi$ and $W$ are labeled by three positive roots $\alpha_1, \alpha_2, \alpha_3$. The root space has been spanned by two simple positive roots. In this case, the free field $\vec{\phi}$ has two components $\phi_1$ and $\phi_2$. Moreover, affine Lie algebra has infinitely many generators or currents, but for now we can consider 8 generators, and then one can expand all these generators by modes. Then, the bosonization of the currents in terms of free fields \cite{Morsl3,GMMOS}:
\begin{equation}
\begin{aligned}
J_{\alpha_1} &= W_1 + \chi_3\,W_2,\\ J_{\alpha_2} &= W_2,\\ J_{\alpha_3} &= W_3,\\
J_{-\alpha_1} &= -\chi_1^{2} W_1 + \chi_2 W_3 - (2-q^{2})\,\partial \chi_1
+ i q\, \chi_1\, \alpha_{1}\!\cdot\!\partial\boldsymbol{\phi},\\
J_{-\alpha_2} &= -\chi_1 \chi_2 W_1 - \chi_2^{2} W_2 - \chi_2 \chi_3 W_3 + \chi_1^{2} \chi_3 W_1
+ (2-q^{2})\, \chi_3\,\partial \chi_1 \\
&\quad - (3-q^{2})\,\partial \chi_2
- i q\, \chi_1 \chi_3\, \alpha_{1}\!\cdot\!\partial\boldsymbol{\phi}
+ i q\, \chi_2\, \alpha_{2}\!\cdot\!\partial\boldsymbol{\phi},\\
J_{-\alpha_3} &=
\chi_1 \chi_3 W_1 - \chi_2 \chi_3 W_2 - \chi_3^{3} W_3 - \chi_2 W_1
- (3-q^{2})\,\partial x_3
+ i q\, \chi_3\, \alpha_{3}\!\cdot\!\partial\boldsymbol{\phi},\\
H_{\alpha_1} &= -2 \chi_1 W_1 - \chi_2 W_2 + \chi_3 W_3 + i q\, \alpha_{1}\!\cdot\!\partial\boldsymbol{\phi},\\
H_{\alpha_2} &= - \chi_1 W_1 - 2 \chi_2 W_2 - \chi_3 W_3 + i q\, \alpha_{2}\!\cdot\!\partial\boldsymbol{\phi},
\end{aligned}
\end{equation}
We can write down these currents in terms of the free fields $u$ and $v$,
\begin{equation}
    \begin{aligned}
       J_{\alpha_1}(z)
&= -\,i\,\partial v_1\,e^{-u_1 + i v_1}
   \;-\; i\,\partial v_2\,e^{\,u_3 - u_2 + i\,(v_2 - v_3)},\\[4pt]
J_{\alpha_2}(z)
&=-\,i\,\partial v_2\,e^{-u_2 + i v_2},\\[4pt]
J_{\alpha_3}(z)
&=-\,i\,\partial v_3\,e^{-u_3 + i v_3}.\\
J_{-\alpha_1} &= e^{u_1-i v_1}\!\left[-(2-q^2)\,\partial u_1 + i\,(3-q^2)\,\partial v_1 + i q\,\alpha_1\!\cdot\!\partial\boldsymbol{\phi}\right] \;-\; i\,(\partial v_3)\,e^{\,u_2-u_3-i\,(v_2-v_3)}\\
J_{-\alpha_2} &= e^{u_2 - i v_2}\!\left[
-(3-q^2)\,\partial u_2
+ i\,\partial v_1 + i\,\partial v_3
+ i\,(4-q^2)\,\partial v_2
+ i q\,\alpha_2\!\cdot\!\partial\boldsymbol{\phi}
\right] \\
&\quad + e^{u_1 + u_3 - i (v_1 + v_3)}\!\left[
(2-q^2)\,\partial u_1
- i\,(3-q^2)\,\partial v_1
- i q\,\alpha_1\!\cdot\!\partial\boldsymbol{\phi}
\right], \\
J_{-\alpha_3} &=e^{u_3 - i v_3}\!\left[ - (3-q^{2})\,\partial u_3
+ i\left( -\partial v_1 + \partial v_2 + (3-q^{2})\,\partial v_3 \right)
+ i q\,\alpha_{3}\!\cdot\!\partial\boldsymbol{\phi} \right] \\
&\qquad + i\,(\partial v_3)\,e^{2u_3 - i\,2 v_3}
+ i\,(\partial v_1)\,e^{u_2 - u_1 - i\,(v_2 - v_1)} \\
H_{\alpha_1} &= i\big( 2\,\partial v_1 + \partial v_2 - \partial v_3 \big)
  + i q\, \alpha_{1}\!\cdot\!\partial\boldsymbol{\phi}, \\ H_{\alpha_2} &= i\big( \partial v_1 + 2\,\partial v_2 + \partial v_3 \big)
  + i q\, \alpha_{2}\!\cdot\!\partial\boldsymbol{\phi}
    \end{aligned}
\end{equation}
The next step is to write an ansatz for the vertex operator as we did in the previous section. We use the analog of (\ref{299}) here. We noticed that in the previous case, there was only one simple positive root and a $\phi$ field was enough. But now, as the rank of the algebra is 2, so we need the field also to be two-component. Namely, we write the ansatz as follows
\begin{equation}
    V(z) = \exp \left( i\lambda \phi(z) + \sum_{\alpha \in \Delta_+} (\sigma_\alpha u_{\alpha}(z) + i\eta_{\alpha} v_{\alpha})(z) \right)
\end{equation}
where $\lambda$, $\sigma_{\alpha}$ and $\eta_{\alpha}$ are three parameters that we need to find. Now, the OPE of this ansatz with the generators

\begin{equation*}
\begin{aligned}
J_{\alpha_2}(w)\,V(z)
&\sim -\frac{\eta_2}{(w-z)^{1-\sigma_2-\eta_2}} \ :e^{-u_2(z)+i v_2(z)}\,V(z):\\[4pt]
J_{\alpha_3}(w)\,V(z)
&\sim -\frac{\eta_3}{(w-z)^{1-\sigma_3-\eta_3}} \ :e^{-u_3(z)+i v_3(z)}\,V(z):\\[4pt]
J_{\alpha_1}(w)\,V(z)
&\sim -\frac{\eta_1}{(w-z)^{1-\sigma_1-\eta_1}} \ :e^{-u_1(z)+i v_1(z)}\,V(z):
      \;- \\& \;\frac{\eta_2}{(w-z)^{1-(\sigma_2-\sigma_3)-(\eta_2-\eta_3)}} \ :e^{\,u_3(z)-u_2(z)+i\,(v_2(z)-v_3(z))}\,V(z):
\end{aligned}
\end{equation*}

\begin{equation}
\begin{aligned}
J_{-\alpha_1}(w)\,V(z)
&\sim \frac{(2-q^2)\sigma_1+(3-q^2)\eta_1+q\,(\alpha_1\!\cdot\!\boldsymbol\lambda)}{(w-z)^{1-\sigma_1-\eta_1}} \ :e^{u_1(z)-i v_1(z)}\,V(z):\\
&\hspace{2.4em}-\frac{\eta_3}{(w-z)^{1-(\sigma_2-\sigma_3)-(\eta_2-\eta_3)}} \ :e^{\,u_2(z)-u_3(z)-i\,(v_2(z)-v_3(z))}\,V(z):,\\[8pt]
J_{-\alpha_2}(w)\,V(z)
&\sim \frac{\eta_1+\eta_2+\eta_3+(3-q^2)(\sigma_2+\eta_2)+q\,(\alpha_2\!\cdot\!\boldsymbol\lambda)}{(w-z)^{1-\sigma_2-\eta_2}} \ :e^{u_2(z)-i v_2(z)}\,V(z):\\
&\hspace{2.4em}-\frac{\,\eta_1+(2-q^2)(\sigma_1+\eta_1)+q\,(\alpha_1\!\cdot\!\boldsymbol\lambda)\,}{(w-z)^{1-(\sigma_1+\eta_1)-(\sigma_3+\eta_3)}} \ :e^{u_1(z)-i v_1(z)}e^{u_3(z)-i v_3(z)}\,V(z):,\\[8pt]
J_{-\alpha_3}(w)\,V(z)
&\sim \frac{-\eta_1+\eta_2+(3-q^2)(\sigma_3+\eta_3)+q\,(\alpha_3\!\cdot\!\boldsymbol\lambda)}{(w-z)^{1-\sigma_3-\eta_3}} \ :e^{u_3(z)-i v_3(z)}\,V(z):\\
&\hspace{2.4em}+\frac{\eta_3}{(w-z)^{1-2(\sigma_3+\eta_3)}} \ :e^{2u_3(z)-2 i v_3(z)}\,V(z):
\;+ \\& \hspace{2.4em} + \;\frac{\eta_1}{(w-z)^{1-(\sigma_1-\sigma_2)-(\eta_1-\eta_2)}} \ :e^{\,u_2(z)-u_1(z)-i\,(v_2(z)-v_1(z))}\,V(z):
\end{aligned}
\end{equation}

\begin{equation*}
\begin{aligned}
H_{\alpha_1}(w)\,V(z)
&\sim \frac{\,2\eta_1+\eta_2-\eta_3+q\,(\alpha_1\!\cdot\!\boldsymbol\lambda)\,}{w-z}\,V(z),\\[4pt]
H_{\alpha_2}(w)\,V(z)
&\sim \frac{\,\eta_1+2\eta_2+\eta_3+q\,(\alpha_2\!\cdot\!\boldsymbol\lambda)\,}{w-z}\,V(z).
\end{aligned}
\end{equation*}
As we have discussed before, the OPE of $V(z)$ with the raising and lowering operators should have a pole of first order and with the Cartan operator gives the residue of the highest weight. These two conditions give the following form of the vertex operator for the highest weight state.
\begin{equation}
    V_{\Lambda} = \exp\left( i\frac{\Lambda}{q} \bm\phi(z) \right)
\end{equation}
Here $\Lambda$ is the highest weight of the algebra, which plays the role of $j$ in the previous section. Now, if we act with the lowering operators, we get the vertex operators corresponding to different weights where the $\chi$ field will appear. We can illustrate this by looking at the simplest example of fundamental and anti-fundamental representation, which is three dimensional. Here, the fundamental weight is $(1,0)$ and anti-fundamental weight is $(0,1)$ which are also the highest weight of this representation. We have seen in the case of $\hat{sl}(2)_k$, all vertex operators were labeled $(j,m)$. In a general sense, for any algebra, this spin-$j$ is actually the highest weight of the representation which we denote by $\Lambda$ here and $m$ is any other weight of that representation which we denote by $\mu$. \\
\newline
Now for any $\Lambda$, we can get all the possible $\mu$ by using the Weyl reflection. For any root $\alpha$, the Cartan matrix is denoted as follows, where all the elements are scalar products.
\begin{equation}
    A_{ji} = \frac{2 (\alpha_i, \alpha_j)}{\alpha_j,\alpha_j}
\end{equation}
Weyl reflection is defined as
\begin{equation}
    s_{\alpha} (\mu) = \mu - 2\frac{(\mu,\alpha)}{(\alpha, \alpha)} \alpha
\end{equation}
The fundamental representation is three dimensional where $\Lambda-$ is the highest weight and $\Lambda-\alpha_1$ and $\Lambda-\alpha_1 - \alpha_2$ are the lower weights, which can be found from the Weyl reflection above. This can also be generalized by the weight diagrams. As we desired to explicitly calculate the correlator, we need explicit expression for the vertex operators, this requires the numerical treatments of the weights and roots. Here we illustrate the weight diagram in Figure \ref{fig:sl3-both} for the fundamental and anti-fundamental representation of $sl(3)$. \\

From now on, we label the vertex operators by $(\Lambda,\mu)$ where both have two components. For the fundamental representation $\Lambda = (1,0)$, the allowed $\mu$ are $\mu \in (1,0), (-1,1),(0,-1)$. If we consider the adjoint representation $\Lambda = (1,1)$ , then it will be 8-dimensional. In the weight diagram, we will get 6 points which create a hexagon and 2 points at the center.\\
\newline
So we now understand how to label the vertex operators. The idea is the same for higher algebras as well. Then the next step is to derive a general expression for the vertex operator and it's reflected/dual version as well.

\subsection{Calculating correlation functions in $\hat{sl}(3)_k$ WZW model}
In this step, we need to express the vertex operator in terms of free fields. In case of $\hat{sl}(2)_k$, we used two types of such operators. One was single $V$-operator and the other was it's reflected version $\tilde{V}$. Now we can also use the same concept here. One of the types of operator corresponds to the fundamental representation $ (V_{\Lambda,\mu})$, another corresponds to anti-fundamental $(V_{\Lambda, \mu}^*)$, and finally the reflected/dual version of them $(\tilde{V}_{\Lambda,\mu})$ and  $(\tilde{V}_{\Lambda,\mu}^*)$. To build the representation of bosonized algebra in the case of the fundamental highest weight $\Lambda_1 = (1,0)$, we proceed with the OPE.
\begin{equation}
    \begin{aligned}
        & V_{\Lambda_1, \mu_1} = \exp \left( i \frac{\Lambda_1}{q} \bm\phi(z) \right) = \exp\left(i\frac{\phi_1}{q}\right); \ \mu_1 = \Lambda_1 = (1,0) \\&
        J_{-\alpha_1} \cdot V_{\Lambda_1, \mu_1} = iq \chi_1 \alpha_1 \partial\bm\phi(w) \cdot \exp\left( i\frac{\Lambda_1}{q} \bm\phi(z)\right) = \frac{1}{w-z} \chi_1 V_{\Lambda_1,\mu_1} = \frac{1}{w-z} V_{\Lambda_1, \mu_2}; \ \mu_2 = \Lambda_1-\alpha_1 \\&
   J_{-\alpha_2} \cdot V_{\Lambda_1,\mu_2} = - \chi_1 \chi_2 W_1 \cdot \chi_1 V_{\Lambda_1} + \chi_1^2 \chi_3 W_1 \cdot \chi_1 V_{\Lambda_1} - iq\chi_1 \chi_3 \alpha_1 \partial \bm\phi\cdot \chi_1 V_{\Lambda_1}  = \\& =  \frac{1}{w-z} \left( - \chi_1\chi_2 V_{\Lambda_1,\mu_1} + \chi_1^2\chi_3 V_{\Lambda_1,\mu_1} - \chi_1^2\chi_3V_{\Lambda_1,\mu_1}  \right) = -\frac{1}{w-z} V_{\Lambda_1,\mu_3}; \ \mu_3 = \Lambda_1-\alpha_1-\alpha_2
    \end{aligned}
\end{equation}
OPE in case of the anti-fundamental weight $\Lambda_2 = (0,1)$:
\begin{equation}
    \begin{aligned}
        & V_{\Lambda_2, \mu_1^*} = \exp \left( i \frac{\Lambda_2}{q} \bm\phi(z) \right); \ \mu_1^* = \Lambda_2 = (0,1) \\&
        J_{-\alpha_2} \cdot V_{\Lambda_2, \mu_1^*} = iq \chi_2 \alpha_2 \partial\bm\phi(w) \cdot \exp\left( i\frac{\Lambda_2}{q} \bm\phi(z)\right) = \frac{1}{w-z} \chi_2 V_{\Lambda_2,\mu_1^*} = \frac{1}{w-z} V_{\Lambda_2, \mu_2^*}; \ \mu_2^* = \Lambda_2-\alpha_2 \\&
   J_{-\alpha_1} \cdot V_{\Lambda_2,\mu_2^*} = 0 \ (\text{no singular contraction terms, so we act} \  V_{\Lambda_2} \ \text{by} \ J_{-\alpha_3}) \\&
   J_{-\alpha_3} V_{\Lambda_2, \mu_1^*} = iq\chi_3 \alpha_3 \partial \bm\phi (\omega) \cdot \exp\left( i\frac{\Lambda_2}{q} \bm\phi(z)\right) =  \frac{1}{w-z} \chi_3 V_{\Lambda_2, \mu_1^*} = \frac{1}{w-z} V_{\Lambda_2, \mu_3^*}; \mu_3^* = \Lambda_2 - \alpha_2 - \alpha_1
    \end{aligned}
\end{equation}
So, finally for the fundamental and anti-fundamental highest weight, we found the representations:
\begin{equation}
  \begin{aligned}
     &  \Lambda_1: \underbrace{V_{\Lambda_1, \mu_1}}_{\Lambda_1}, \ \underbrace{ \chi_1V_{\Lambda_1, \mu_1}}_{\Lambda_1 - \alpha_1}, \ \underbrace{\chi_1\chi_2V_{\Lambda_1, \mu_1}}_{\Lambda_1 - \alpha_1- \alpha_2}\\& \Lambda_2: \underbrace{V_{\Lambda_2, \mu_1^*}}_{\Lambda_2}, \ \underbrace{ \chi_2V_{\Lambda_2, \mu_1^*}}_{\Lambda_2 - \alpha_2}, \ \underbrace{\chi_3 V_{\Lambda_2, \mu_1^*}}_{\Lambda_2 - \alpha_2 - \alpha_1}
  \end{aligned}
\end{equation}
Now, the idea is the same to build any representation of the bosonized algebra. The procedure is also the same for higher algebra as well. Due to the dependence on the metric of the sphere, we introduce the vacuum charge at point $R$:
\begin{equation}
    V_s(R) = \exp\left( \frac{i}{q} (2\phi_1(R)+ 2\phi_2(R) + u_1+u_2+u_3 - i(v_1+v_2+v_3))\right)
\end{equation}
Now we can notice that inserting the vertex operators (and their duals) above with the vacuum charge will not fulfill the charge neutrality condition. Similarly as in the case of $\hat{sl}(2)_k$, here we need to insert at least one reflected/conjugate vertex operator that has the same conformal dimension but with negative signs in fields to dismiss the remaining free fields in the exponent.
\begin{equation}
    \tilde{V}_{\Lambda_2, \mu_1^*}^* = \exp{\left( -\frac{i}{q} (3\Lambda_2 +2\Lambda_1 ) \bm\phi -u_1-u_2-u_3+i(v_1+v_2+v_3)\right)}
\end{equation}
Now, we are ready to calculate the two-point function.
\begin{equation}
\begin{aligned}
  & \Bigr \langle V_{\Lambda_1, \mu_1}(w) \tilde{V}_{\Lambda_2, \mu_1^*}^*(z)  V_s(R) \Bigr \rangle = \Bigr \langle V_{(1,0),(1,0) }(w) \tilde{V}_{(0,1) (0,1)}^* (z) V_s(R) \Bigr \rangle = \\& \Bigr \langle e^{\frac{i}{q} \phi_1(w)} e^{-\frac{i}{q} (3\phi_1 + 2\phi_2) -u_1-u_2-u_3+i(v_1+v_2+v_3)} e^{\frac{i}{q} (2\phi_1(R)+ 2\phi_2(R) + u_1+u_2+u_3 - i(v_1+v_2+v_3))}  \Bigr \rangle  = \frac{1}{(w-z)^{2\Delta}}
   \end{aligned}
\end{equation}
where \begin{equation}
    \Delta = \frac{1}{2q^2} \Lambda_1 (3\Lambda_1 + 2\Lambda_2) = \frac{1}{2q^2} \Lambda_2 (3\Lambda_2 + 2\Lambda_1)
\end{equation}
We can see that the charge neutrality condition holds and that the correlator depends on the conformal dimension of the operator and that no screening charge is required. Now, let us consider the following four-point function. For now, we consider only the fundamental and anti-fundamental highest weight.
\begin{equation}
    \Bigr \langle \tilde{V}_{(0,1), (-1,0)}^* (0) V_{(1,0),(1,0)} (1) V_{(0,1),(-1,0)}^*(x) V_{(1,0),(1,0)}(\infty) V_s(R) \Bigr \rangle
    \label{10390}
\end{equation}
Let's write the explicit expressions of these operators in terms of free fields ($\phi,u,v$). We get
\begin{equation}
    \begin{aligned}
         &\tilde{V}_{(0,1), (-1,0)}^* (0) = \exp \left(-\frac{i}{q} (3\phi_1 (0) + 2\phi_2 (0)) -u_1(0)-u_2(0)-u_3(0)+i(v_1(0)+v_2(0)+v_3(0))\right) \\&
         V_{(1,0),(1,0)} (1) = \exp \left( \frac{i}{q} \phi_1 (1) \right) \\&
          V_{(0,1),(-1,0)}^*(x) = \exp \left( \frac{i \phi_2 (x)}{q} + u_2 (x) - iv_2 (x) \right) \\&
          V_{(1,0),(1,0)}(\infty) = \exp \left( \frac{i \phi_1 (\infty)}{q} \right)
    \end{aligned}
\end{equation}
In this case, we see that the charge neutrality condition in (\ref{10390}) doesn't hold. And we need to insert screening operators. Here we have two types of screening charges due to the rank of the algebra is $2$. We have
\begin{equation}
    \begin{aligned}
        &Q_1 = \oint W_1 \exp\left( -\frac{i}{q} \alpha_1 \bm \phi\right) = \oint \exp \left( -\frac{2i \phi_1}{q} + \frac{i\phi_2}{q} -u_1 + iv_1  \right) \frac{1}{i} \partial v_1 \\& Q_2 = \oint \chi_1W_2 \exp \left( -\frac{i}{q} \alpha_2 \bm\phi \right) = \oint \exp \left( \frac{i\phi_1}{q} - \frac{2i\phi_2}{q}  + u_1 - u_2 -i (v_1 - v_2)\right) \frac{1}{i} \partial v_2
    \end{aligned}
\end{equation}
 The correlator now have the following form
\begin{equation}
      \Bigr \langle \tilde{V}_{(0,1), (-1,0)}^* (0) V_{(1,0),(1,0)} (1) V_{(0,1),(-1,0)}^*(x) V_{(1,0),(1,0)}(\infty) V_s(R) Q_1(t_1) Q_2(t_2)\Bigr \rangle
\end{equation}
We can check the charge neutrality condition of each fields at those four points
\begin{equation}
    \begin{aligned}
       & \phi_1 : \ -\frac{3i}{q} + \frac{i}{q}+ \frac{i}{q}- \frac{2i}{q}+ \frac{i}{q} + \frac{2i }{ q} = 0 \\&
       \phi_2: \  -2 + \frac{i}{q} + \frac{i}{q} - \frac{2i}{q} +2  = 0 \\&
       u_1: \ -1-1+1+1 = 0 \\&
       u_2: \ -1+1-1+1 = 0 \\&
       u_3: \ -1+1 = 0 \\&
       v_1: \ i+i-i-i = 0 \\&
       v_2: \ i-i+i-i=0 \\&
       v_3: \ i-i = 0
    \end{aligned}
\end{equation}
Next, we can check the tensor product of the highest weights in irreducible decomposition
\begin{equation}
    (0,1)\otimes(1,0)\otimes(0,1)\otimes(1,0)
= 2\,(0,0)\ \oplus\ 4\,(1,1)\ \oplus\ (3,0)\ \oplus\ (0,3)\ \oplus\ (2,2)
\end{equation}
We find that the chosen highest weights produce two singlets. Now to proceed to the calculation of the correlator, the contractions of the exponential parts that take part in the integral,
\begin{equation}
    \begin{aligned}
      & t_1^{- \left( -\frac{2i}{q} \right)  \left( -\frac{3i}{q} \right) -  \left( \frac{i}{q}  \left( -\frac{2i}{q} \right) -  \right) 1 +1 } = t_1^{-\frac{4}{q^2} } \\& t_2^{-  \left( \frac{i}{q} \right)  \left( -\frac{3i}{q} \right) -  \left( -\frac{2i}{q} \right) \left( -\frac{2i}{q} \right) + 1- 1-1+1} =  t_2^{ \frac{1}{q^2} } \\& (1-t_1)^{ - \left( \frac{i}{q} \right)  \left( -\frac{2i}{q} \right)} = (1-t_1)^{ -\frac{2}{q^2}} \\& (1-t_2)^{  - \left( \frac{i}{q} \right)  \left( \frac{i}{q} \right) } = (1-t_2)^{\frac{1}{q^2}} \\& (t_1-x)^{-  \left( \frac{i}{q} \right)  \left( \frac{i}{q} \right) } = (t_1- x)^{\frac{1}{q^2}} \\& (t_2 -x)^{ - \left(- \frac{2i}{q} \right)  \left( \frac{i}{q} \right)} = (t_2 - x)^{\frac{2}{q^2}} \\& (t_1 - t_2)^{ - \left( \frac{i}{q} \right)  \left( -  \frac{2i}{q} \right) - \left( -\frac{2i}{q} \right)  \left( \frac{i}{q} \right) +1-1} = (t_1-t_2)^{-\frac{4}{q^2}}
    \end{aligned}
\end{equation}
all together
\begin{equation}
    t_1^{\frac{4}{q^{2}}}\, t_2^{\frac{1}{q^{2}}}\,
(1-t_1)^{-\frac{1}{q^{2}}}\,
(1-t_2)^{\frac{1}{q^{2}}}\,
(t_1-x)^{\frac{1}{q^{2}}}\,
(t_2-x)^{-\frac{2}{q^{2}}}\,
(t_1-t_2)^{-\frac{4}{q^{2}}}
\end{equation}
To find the contractions with the differential part ($\partial v_1$ and $\partial v_2$), we use (\ref{diff contraction}). The possible contractions are
\begin{equation}
    \begin{aligned}
        & \frac{1}{i} c_k \Bigr \langle \partial v_1 (t_1) v_1 (0)\Bigr \rangle = \frac{1}{i} \left( -\frac{i}{t_1} = \right) = - \frac{1}{t_1} \\& \frac{1}{i} c_k \Bigr \langle \partial v_1 (t_1) v_1 (t_2)\Bigr \rangle = \frac{1}{i} (-i) \left( -\frac{1}{t_1 - t_2} \right) = \frac{1}{t_1 - t_2} \\& \frac{1}{i} c_k \Bigr \langle \partial v_2 (t_2) v_2 (0)\Bigr \rangle = \frac{1}{i} \left( -\frac{i}{t_2} = \right) = - \frac{1}{t_2} \\& \frac{1}{i} c_k \Bigr \langle \partial v_2 (t_2) v_2 (x)\Bigr \rangle = \frac{1}{i} (-i) \left( -\frac{1}{t_2 - x} \right) = \frac{1}{t_2-x}
    \end{aligned}
\end{equation}
all together
\begin{equation}
    \left(  \frac{1}{t_1} - \frac{1}{t_1 - t_2}\right) \left( \frac{1}{t_2} - \frac{1}{t_2 - x} \right) = \frac{x}{t_1 (t_2-x) (t_1-t_2)}
\end{equation}
Finally, the correlator
\begin{equation}
\begin{aligned}
   & F \ = \ \Bigr \langle \tilde{V}_{(0,1), (-1,0)}^* (0) V_{(1,0),(1,0)} (1) V_{(0,1),(-1,0)}^*(x) V_{(1,0),(1,0)}(\infty) V_s(R) Q_1(t_1) Q_2(t_2) \Bigr \rangle  =\\&   \boxed{ x^{-\frac{2}{q^2}}  \oint \oint dt_1 dt_2 \left[ t_1^{-1+\frac{4}{q^2}} t_2^{\frac{1}{q^2}}(1-t_1)^{-\frac{2}{q^2}} (1-t_2)^{\frac{1}{q^2}} (t_1 - x)^{\frac{1}{q^2}} (t_2 -x)^{-1-\frac{2}{q^2}}(t_1 - t_2)^{-1-\frac{4}{q^2}}  \right]}
   \label{KZ2}
\end{aligned}
\end{equation}
This is the DF integral representation of the four-point correlator in the $\hat{sl}(3)_k$ WZW model, where the vertex operators are labeled by two fundamental and anti-fundamental representations.

\newpage
\subsection{Solving Knizhnik-Zamolodchikov equations}
This particular integral form of the correlator (\ref{KZ2}) solves the Knizhnik-Zamoldochikov (KZ) equation (\ref{kz}). At these fixed four points ($0,1,x,\infty$), the KZ equation takes the following form
\begin{equation}
    (k+3) \ \partial_x F  - \left( \frac{ \sum_a t_1^a \otimes t_2^a}{x} + \frac{ \sum_a t_1^a \otimes t_3^a}{x-1} \right) F = 0
    \label{kz3}
\end{equation}
In case of $\hat{sl}(2)_k$, we show that the correlator is $SU(2)$ invariant. In the case of $\hat{sl}(3)_k$, we act on the correlator by the generator of $SU(3)$. In our case, the DF representation of the correlator (\ref{KZ2}) is actually a sum of conformal blocks ($F_i$). If we consider the vertex operator corresponding to some higher highest weight, the number of conformal blocks will be different. To show the equivalence of the KZ equation with the DF integral, we act on the integral by the generators of $SU(3)$. The generator $t_1^a$ acts on the vertex operator corresponding to point $x$, $t_2^a$ acts on the vertex operator corresponding to point $0$ and $t_3^a$ on  $1$, which can be explicitly formulated from the Gell-Mann matrices. In our specific example (\ref{KZ2}), we considered two types of vertex operators corresponding to the fundamental highest weight and the anti-fundamental highest weight. So let us see how the $8$ generators of $SU(3$ act on them
\begin{equation}
    \begin{aligned}
        t^1 V_{(1,0),(1,0)} =& \frac{1}{2}V_{(1,0), (-1,1)}; \ \ \ \ t^1 V_{(0,1),(-1,0)} = -\frac{1}{2}V_{(0,1), (1,-1)}; \\
         t^2 V_{(1,0),(1,0)} =& \frac{i}{2}V_{(1,0), (-1,1)}; \ \ \ \ t^2 V_{(0,1),(-1,0)} = \frac{i}{2}V_{(0,1), (1,-1)}; \\
          t^3 V_{(1,0),(1,0)} =& \frac{1}{2}V_{(1,0), (1,0)}; \ \ \ \ t^1 V_{(0,1),(-1,0)} = -\frac{1}{2}V_{(0,1), (-1,0)}; \\ t^4 V_{(1,0),(1,0)} =& \frac{1}{2}V_{(1,0), (0,-1)}; \ \ \ \ t^4 V_{(0,1),(-1,0)} = -\frac{1}{2}V_{(0,1), (0,1)}; \\ t^5 V_{(1,0),(1,0)} =& \frac{i}{2}V_{(1,0), (0,-1)}; \ \ \ \ t^5 V_{(0,1),(-1,0)} = \frac{i}{2}V_{(0,1), (0,1)}; \\ t^6 V_{(1,0),(1,0)} =& \ t^7 V_{(1,0),(1,0)} = 0; \ \ \ \ t^6 V_{(0,1),(-1,0)} = t^7 V_{(0,1),(-1,0)} =  0
           \\ t^8 V_{(1,0),(1,0)} =& \frac{1}{2\sqrt{3}} V_{(1,0),(1,0)}; \ \ \ \ t^8 V_{(0,1),(-1,0)} = -\frac{1}{2\sqrt{3}} V_{(0,1),(-1,0)}
    \end{aligned}
\end{equation}
Next, we calcuate the action of $\sum_a t_1^a \otimes t_2^a$ on the correlator.
\begin{equation}
    \begin{aligned}
       &  \left(\sum_{a=1}^8 t_1^a \otimes t_2^a\right) F = \\& =  \left( t_1^1 \otimes t_2^1 \right) F + \left( t_1^2 \otimes t_2^2 \right) F+ \left( t_1^3 \otimes t_2^3 \right) F + \left( t_1^4 \otimes t_2^4 \right) F + \left( t_1^5 \otimes t_2^5 \right) F + \left( t_1^6 \otimes t_2^6 \right) F + \left( t_1^7 \otimes t_2^7 \right) F + \left( t_1^8 \otimes t_2^8 \right) F =\\& = \frac{1}{4} x^{-\frac{2}{q^2}}  \underbrace{\Bigr \langle \tilde{V}_{(0,1), (1,-1)}^* (0) V_{(1,0),(1,0)} (1) V_{(0,1),(1,-1)}^*(x) V_{(1,0),(1,0)}(\infty) V_s(R) Q_1(t_1) Q_2(t_2) \Bigr \rangle}_{= \ 0}  - \\& -  \frac{1}{4} x^{-\frac{2}{q^2}}  \underbrace{\Bigr \langle \tilde{V}_{(0,1), (1,-1)}^* (0) V_{(1,0),(1,0)} (1) V_{(0,1),(1,-1)}^*(x) V_{(1,0),(1,0)}(\infty) V_s(R) Q_1(t_1) Q_2(t_2) \Bigr \rangle}_{= \ 0} + \\& + \frac{1}{4} x^{-\frac{2}{q^2}}  \Bigr \langle \tilde{V}_{(0,1), (-1,0)}^* (0) V_{(1,0),(1,0)} (1) V_{(0,1),(-1,0)}^*(x) V_{(1,0),(1,0)}(\infty) V_s(R) Q_1(t_1) Q_2(t_2) \Bigr \rangle + \\& +
       \frac{1}{4} x^{-\frac{2}{q^2}}  \underbrace{\Bigr \langle \tilde{V}_{(0,1), (0,1)}^* (0) V_{(1,0),(1,0)} (1) V_{(0,1),(0,1)}^*(x) V_{(1,0),(1,0)}(\infty) V_s(R) Q_1(t_1) Q_2(t_2) \Bigr \rangle}_{= \ 0} - \\& -  \frac{1}{4} x^{-\frac{2}{q^2}}  \underbrace{\Bigr \langle \tilde{V}_{(0,1), (0,1)}^* (0) V_{(1,0),(1,0)} (1) V_{(0,1),(0,1)}^*(x) V_{(1,0),(1,0)}(\infty) V_s(R) Q_1(t_1) Q_2(t_2) \Bigr \rangle}_{= \ 0}  + \\& +
        \frac{1}{12} x^{-\frac{2}{q^2}}  \Bigr \langle \tilde{V}_{(0,1), (-1,0)}^* (0) V_{(1,0),(1,0)} (1) V_{(0,1),(-1,0)}^*(x) V_{(1,0),(1,0)}(\infty) V_s(R) Q_1(t_1) Q_2(t_2) \Bigr \rangle = \\& =  \frac{1}{3} x^{-\frac{2}{q^2}}  \Bigr \langle \tilde{V}_{(0,1), (-1,0)}^* (0) V_{(1,0),(1,0)} (1) V_{(0,1),(-1,0)}^*(x) V_{(1,0),(1,0)}(\infty) V_s(R) Q_1(t_1) Q_2(t_2) \Bigr \rangle = \\& = \frac{1}{3} x^{-\frac{2}{q^2}} F
    \end{aligned}
\end{equation}
The same we do now for the generator acting at points $x$ and $1$.
\begin{equation}
    \begin{aligned}
       &  \left(\sum_{a=1}^8 t_1^a \otimes t_3^a\right) F = \\& =  \left( t_1^1 \otimes t_3^1 \right) F + \left( t_1^2 \otimes t_3^2 \right) F+ \left( t_1^3 \otimes t_3^3 \right) F + \left( t_1^4 \otimes t_3^4 \right) F + \left( t_1^5 \otimes t_3^5 \right) F + \left( t_1^6 \otimes t_3^6 \right) F + \left( t_1^7 \otimes t_3^7 \right) F + \left( t_1^8 \otimes t_3^8 \right) F =\\& = -\frac{1}{4} x^{-\frac{2}{q^2}}  \Bigr \langle \tilde{V}_{(0,1), (-1,0)}^* (0) V_{(1,0),(-1,1)} (1) V_{(0,1),(1,-1)}^*(x) V_{(1,0),(1,0)}(\infty) V_s(R) Q_1(t_1) Q_2(t_2) \Bigr \rangle - \\& -  \frac{1}{4} x^{-\frac{2}{q^2}}  \Bigr \langle \tilde{V}_{(0,1), (-1,0)}^* (0) V_{(1,0),(-1,1)} (1) V_{(0,1),(1,-1)}^*(x) V_{(1,0),(1,0)}(\infty) V_s(R) Q_1(t_1) Q_2(t_2) \Bigr \rangle + \\& - \frac{1}{4} x^{-\frac{2}{q^2}}  \Bigr \langle \tilde{V}_{(0,1), (-1,0)}^* (0) V_{(1,0),(1,0)} (1) V_{(0,1),(-1,0)}^*(x) V_{(1,0),(1,0)}(\infty) V_s(R) Q_1(t_1) Q_2(t_2) \Bigr \rangle - \\& -
       \frac{1}{4} x^{-\frac{2}{q^2}}  \Bigr \langle \tilde{V}_{(0,1), (-1,0)}^* (0) V_{(1,0),(0,-1)} (1) V_{(0,1),(0,1)}^*(x) V_{(1,0),(1,0)}(\infty) V_s(R) Q_1(t_1) Q_2(t_2) \Bigr \rangle - \\& -  \frac{1}{4} x^{-\frac{2}{q^2}}  \Bigr \langle \tilde{V}_{(0,1), (-1,0)}^* (0) V_{(1,0),(0,-1)} (1) V_{(0,1),(0,1)}^*(x) V_{(1,0),(1,0)}(\infty) V_s(R) Q_1(t_1) Q_2(t_2) \Bigr \rangle  - \\& -
        \frac{1}{12} x^{-\frac{2}{q^2}}  \Bigr \langle \tilde{V}_{(0,1), (-1,0)}^* (0) V_{(1,0),(1,0)} (1) V_{(0,1),(-1,0)}^*(x) V_{(1,0),(1,0)}(\infty) V_s(R) Q_1(t_1) Q_2(t_2) \Bigr \rangle = \\& =  -\frac{4}{3} x^{-\frac{2}{q^2}}  \Bigr \langle \tilde{V}_{(0,1), (-1,0)}^* (0) V_{(1,0),(-1,1)} (1) V_{(0,1),(1,-1)}^*(x) V_{(1,0),(1,0)}(\infty) V_s(R) Q_1(t_1) Q_2(t_2) \Bigr \rangle = \\& = -\frac{4}{3} x^{-\frac{2}{q^2}} F
    \end{aligned}
\end{equation}
Now gather everything together and recall $(k+3) = q^2$
\begin{equation}
    \begin{aligned}
        q^2 \frac{\partial F}{\partial x} &= \left[ \frac{1}{3x} - \frac{4}{3(x-1)} \right] F
    \end{aligned}
    \label{kzkzkz}
\end{equation}
The symmetry  of the correlator simplified the KZ equation. This equation (\ref{kzkzkz}) is actually not a scaler as $F$ can be decomposed into a finite number of conformal blocks. Then this vector equation can be reduced to a scalar equation like (\ref{hypergeo}), as we saw in the previous section.  But for a higher number of conformal blocks, how such a large vector equation can be reduced to a scalar equation of some finite order is not very obvious from the point of view of the free field approach. However, we can try to demonstrate how such combination of conformal blocks solves the original KZ equation using the total derivative technique. \\
\newline
For convenience in writing, let us express the integrand of (\ref{KZ2}) as $$
I(t_1,t_2;x) = x^{-\frac{2}{q^2}} \left[ t_1^{-1+\frac{4}{q^2}} t_2^{\frac{1}{q^2}}(1-t_1)^{-\frac{2}{q^2}} (1-t_2)^{\frac{1}{q^2}} (t_1 - x)^{\frac{1}{q^2}} (t_2 -x)^{-1-\frac{2}{q^2}}(t_1 - t_2)^{-1-\frac{4}{q^2}}  \right]
$$
and then, from (\ref{kzkzkz}), we can write
\begin{equation}
    \begin{aligned}
        & q^2  \oint \oint dt_1 dt_2 I(t_1,t_2;x) \left[ -\frac{1}{q^2(t_1-x)} + \left(1+ \frac{2}{q^2}\right) \frac{1}{t_2-x} - \frac{2}{q^2} \frac{1}{x} \right] =  \oint \oint dt_1 dt_2 I(t_1,t_2;x) \left[ \frac{1}{3x} - \frac{4}{3(x-1)} \right]
    \end{aligned}
\end{equation}
from which
\begin{equation}
       \oint \oint dt_1 dt_2 I(t_1,t_2;x) \underbrace{\left[ -\frac{1}{(t_1-x)} +  \frac{2+q^2}{t_2-x} - \frac{2}{x} - \frac{1}{3x} + \frac{4}{3(x-1)}  \right]}_{J(t_1,t_2;x)} = 0
\end{equation}

Now to hold this, we need to express the whole integrad as a total derivative of $t_1$ and $t_2$. That is, we need to find functions $H(t_1,t_2)$ and $H(t_1,t_2)$ such that
\begin{equation}
   I(t_1,t_2;x) J(t_1,t_2;x) =  \frac{\partial H_1}{\partial t_1} + \frac{\partial H_2}{\partial t_2}
   \label{an11}
\end{equation}
Eventually, we found such functions to be
\begin{equation}
    \begin{aligned}
       & H_1(t_1,t_2) =  I(t_1, t_2;x) \left( \left(-1-\frac{1}{q^2}\right) \ln(t_1-x) + \left( 1-\frac{4}{q^2}\right) \ln t_1+ \left(\frac{2}{q^2}\right) \ln (1-t_1) +  \frac{4 t_1}{3 (1+x)}\right) \\&
       H_2 (t_1,t_2) = I(t_1,t_2;x) \left( \left(3+q^2+\frac{2}{q^2}\right) \ln(t_2-x) - \left(\frac{1}{q^2}\right) \ln t_2- \left(\frac{1}{q^2}\right) \ln (1-t_2) - \frac{7 t_2}{3x} \right)
    \end{aligned}
    \label{an5}
\end{equation}
Let us discuss in short how to find such functions. We use the same analysis as we did in the previous section. Let us
\begin{equation}
\begin{aligned}
   &  H_1 = I(t_1,t_2;x) A(t_1,t_2;x) \\& H_2 = I(t_1,t_2;x) B(t_1,t_2;x)
    \end{aligned}
\end{equation}
Then (\ref{an11}) will be
\begin{equation}
     I(t_1,t_2;x) J(t_1,t_2;x)  = \frac{\partial I}{\partial t_1} A + I \frac{\partial A}{\partial t_1} + \frac{\partial I}{\partial t_2} B + I \frac{\partial B}{\partial t_2}
     \label{an12}
\end{equation}
Then
\begin{equation}
\begin{aligned}
  &  \frac{\partial }{\partial t_1} (\ln I) = \frac{1}{I} \frac{\partial I}{\partial t_1} \Rightarrow \frac{\partial I}{\partial t_1}  =  I \left( \frac{\frac{2}{q^2}}{1-t_1} + \frac{1+\frac{4}{q^2}}{t_1} - \frac{1+\frac{4}{q^2}}{t_1 - t_2} + \frac{\frac{1}{q^2}}{t_1 - x} \right)\\&
  \frac{\partial }{\partial t_2} (\ln I) = \frac{1}{I} \frac{\partial I}{\partial t_2} \Rightarrow \frac{\partial I}{\partial t_2}  =  I \left( -\frac{\frac{1}{q^2}}{1-t_2} + \frac{\frac{1}{q^2}}{t_2} + \frac{1+\frac{4}{q^2}}{t_1 - t_2} - \frac{1+\frac{2}{q^2}}{t_2 - x} \right)
    \end{aligned}
\end{equation}
Then (\ref{an12}) will be
\begin{equation}
     I(t_1,t_2;x)J(t_1,t_2;x) = I(t_1,t_2;x) \underbrace{\left( \frac{\frac{2}{q^2}}{1-t_1} + \frac{1+\frac{4}{q^2}}{t_1} + \frac{\frac{1}{q^2}}{t_1 - x}   -\frac{\frac{1}{q^2}}{1-t_2} + \frac{\frac{1}{q^2}}{t_2} - \frac{1+\frac{2}{q^2}}{t_2 - x} + \frac{\partial A}{\partial t_1} + \frac{\partial B}{\partial t_2}  \right)}_{=J(t_1,t_2;x)}
     \label{an3}
\end{equation}
To obtain $J(t_1,t_2;x)$ in the RHS of (\ref{an3}), we write an ansatz for $A$ and $B$
\begin{equation}
   \begin{aligned}
       & A = a_1 \ln (t_1 -x) + a_2 \ln(t_1) + a_3 \ln (1-t_1) + \mathcal{A}(t_1, x) \\&
    B = b_1 \ln (t_1 -x) + b_2 \ln(t_1) + b_3 \ln (1-t_1) + \mathcal{B}(t_2,x)
   \end{aligned}
   \label{an4}
\end{equation}
After comparing the coefficients, we find
\begin{equation}
\begin{aligned}
    & a_1 = -1-\frac{1}{q^2}; \ a_2 = 1-\frac{4}{q^2}; \ a_3 = \frac{2}{q^2}; \ \mathcal{A}(t_1,x) = \frac{4 t_1}{3(1+x)} \\&
    b_1 = 3+ q^2 +\frac{2}{q^2}; \ b_2 = -\frac{1}{q^2}; \ b_3 = -\frac{1}{q^2};; \ \mathcal{B}(t_2,x) = -\frac{7 t_2}{3 x}
\end{aligned}
\end{equation}
after substituting these coefficients into (\ref{an4}), we obtain (\ref{an5}). \\
\newline
Now, as we discussed above, the contours can be represented as integrals around two cuts — between 0 and 1 and between 1 and $\infty$. And in our case, there are two of such choices, thus the total number is 4.\cite{DF},\cite{DF2} as

\begin{equation}
    \begin{aligned}
       & F_1 = \int_1^{\infty} dt_1 \int_1^\infty dt_2  \left[ t_1^a t_2^b (1-t_1)^c (1-t_2)^d(x-t_1)^e (x-t_2)^f (t_1 - t_2)^g \right] \\&
       F_2 = \int_0^{1} dt_1 \int_1^\infty dt_2  \left[ t_1^a t_2^b (1-t_1)^c (1-t_2)^d(x-t_1)^e (x-t_2)^f (t_1 - t_2)^g \right]  \\&
        F_3= \int_1^{\infty} dt_1 \int_0^1 dt_2  \left[ t_1^a t_2^b (1-t_1)^c (1-t_2)^d(x-t_1)^e (x-t_2)^f (t_1 - t_2)^g \right] \\&
       F_4= \int_0^{1} dt_1 \int_0^1 dt_2  \left[ t_1^a t_2^b (1-t_1)^c (1-t_2)^d(x-t_1)^e (x-t_2)^f (t_1 - t_2)^g \right]
    \end{aligned}
\end{equation}
with
\begin{equation}
    a = -1+\frac{4}{q^2}; \ \ b = d = e = \frac{1}{q^2}; \ \ c=-\frac{2}{q^2}; \ \ f=-1-\frac{2}{q^2}; \ \ g=-1-\frac{4}{q^2}
\end{equation}
This suggests that the integrals should satisfy a fourth-order differential equation. One should be able to extract such a scalar equation from the multi-component KZ equations for $\hat{sl}(3)_k$ model. 

\newpage
\section{Towards the connection of multipoint correlator, higher algebra, and representations}
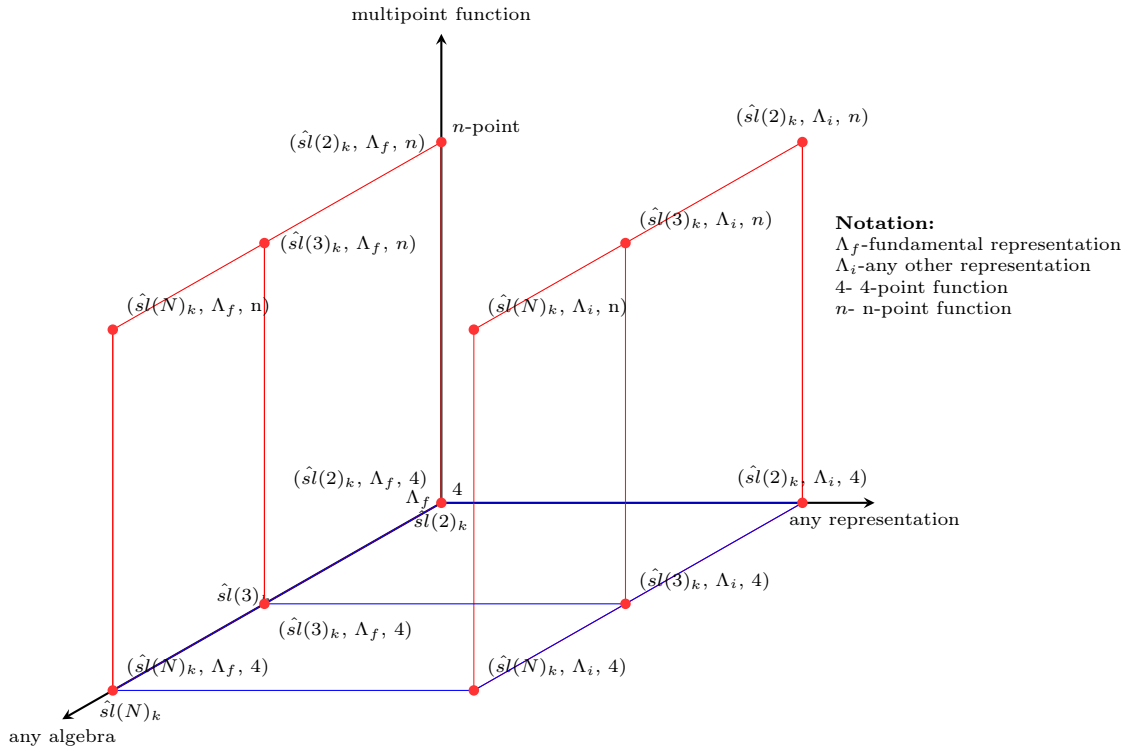
\begin{figure}[H]
\centering
\begin{tikzpicture}[
  x={(-0.7cm,-0.4cm)},
  y={(1cm,0cm)},
  z={(0cm,1cm)},
  scale=2.4,
  >=stealth,
  axis/.style={->,thick},
  dot/.style={circle,fill=black,inner sep=1.2pt},
  highlight/.style={circle,fill=red!80,inner sep=1.4pt},
  labelstyle/.style={font=\scriptsize}
]

\coordinate (O) at (0,0,0);
\draw[axis] (O) -- (3.0,0,0) node[below, labelstyle]{any algebra};
\draw[axis] (O) -- (0,2.4,0) node[below, labelstyle]{any representation};
\draw[axis] (O) -- (0,0,2.6) node[above, labelstyle]{multipoint function};

\node[labelstyle] at (0,0,-0.1) {$\hat{sl}(2)_k$};
\node[labelstyle] at (0,1, -0.25) {};
\node[labelstyle] at (0,2, -0.25) {};

\node[labelstyle] at (0.2,0.14,0.1) [left] {$\Lambda_f$};
\node[labelstyle] at (-0.7 , 0, 1.8) [left] {$n$-point};

\node[labelstyle] at (0.1,0.16,0.2) [below] {4};
\node[labelstyle] at (0.99,-0.4,0) [below] {$\hat{sl}(3)_k $};
\node[labelstyle] at (2.6,0.1,0) [below] {$\hat{sl}(N)_k $};

\foreach \x in {0,1.4,2.6}{
  \draw[red!90] (\x,0,0) -- (\x,0,2);
  \draw[red!90] (\x,2,0) -- (\x,2,2);
}

\foreach \z in {0,2}{
  \draw[red!100] (0,0,\z) -- (2.6,0,\z);
  \draw[red!100] (0,2,\z) -- (2.6,2,\z);
}

\foreach \x in {0,1.4,2.6}{
  \draw[blue!100] (\x,0,0) -- (\x,2,0);
  \draw[red!100] (\x,0,0) -- (\x,0,2);
}
\foreach \y in {0,2}{
  \draw[blue!100] (0,\y,0) -- (2.6,\y,0);
  \draw[red!100] (0,\y,0) -- (0,\y,2);
}

\node[highlight, label={[labelstyle]above left:{ ($\hat{sl}(2)_k$, $\Lambda_f$, 4)}}] at (0,0,0) {};

\node[highlight, label={[labelstyle]left:{$(\hat{sl}(2)_k$, $\Lambda_f$, $n$)}}] at (0,0,2) {};

\node[highlight, label={[labelstyle]above:{($\hat{sl}(2)_k$, $\Lambda_i$, 4})}] at (0,2,0) {};
\node[highlight, label={[labelstyle]above:{($\hat{sl}(2)_k$, $\Lambda_i$, $n$)}}] at (0,2,2) {};

\node[highlight, label={[labelstyle]below right:{($\hat{sl}(3)_k$, $\Lambda_f$, 4})}] at (1.4,0,0) {};
\node[highlight, label={[labelstyle]right:{($\hat{sl}(3)_k$, $\Lambda_f$, $n$})}]   at (1.4,0,2) {};

\node[highlight, label={[labelstyle]above right:{($\hat{sl}(3)_k$, $\Lambda_i$, 4)}}] at (1.4,2,0) {};
\node[highlight, label={[labelstyle]above right:{$(\hat{sl}(3)_k$, $\Lambda_i$, $n$)}}] at (1.4,2,2) {};

\node[highlight, label={[labelstyle]above right:{($\hat{sl}(N)_k$, $\Lambda_f$, 4)}}] at (2.6,0,0) {};
\node[highlight, label={[labelstyle]above right:{($\hat{sl}(N)_k$, $\Lambda_i$, 4)}}] at (2.6,2,0) {};
\node[highlight, label={[labelstyle]above right:{($\hat{sl}(N)_k$, $\Lambda_f$, n)}}] at (2.6,0,2) {};
\node[highlight, label={[labelstyle]above right:{($\hat{sl}(N)_k$, $\Lambda_i$, n)}}] at (2.6, 2, 2) {};

\node[labelstyle, align=left, anchor=west] at (7.1,7.1,4.1) {
  \textbf{Notation:}\\
  $\Lambda_f$-fundamental representation\\
   $\Lambda_i$-any other representation \\
  $4$- 4-point function \\
   $n$- n-point function \\
};
\end{tikzpicture}
\caption{Connection between multipoint correlator, higher algebra, and representations }
\label{fig22}
\end{figure}

To begin with, let us consider the following example. We are interested in the example of such 6-point correlator in $\hat{sl}(2)_k$ WZW model with all fundamental representations as follows:
\begin{equation}
\Lambda_1 = \Lambda_2 = \Lambda_3 = \Lambda_4 =  \Lambda_5 = \Lambda_6 = \frac{1}{2}
\end{equation}
The correlator in this case
\begin{equation}
     \Bigl \langle \tilde{V}_{\frac{1}{2},-\frac{1}{2}}(0) V_{\frac{1}{2}, \frac{1}{2}} (x_1) V_{\frac{1}{2}, \frac{1}{2}} (x_2) V_{\frac{1}{2}, -\frac{1}{2}} (x_3)  V_{\frac{1}{2},\frac{1}{2}} (1) V_{\frac{1}{2},-\frac{1}{2}}(\infty)   V_s(\infty) Q_1 (t_1) Q_2(t_2) \Bigr \rangle
\end{equation}
The corresponding DF integral for this correlator will have two screenings, means have two contour integrals. Then it will take the following form
\begin{equation}
  F=  \oint dt_1 \oint dt_2 [ t_1^{a_1} t_2^{a_2} (1-t_1)^{a_3} (1-t_2)^{a_4} (t_1 - x_1)^{a_5} (t_1 - x_2)^{a_6} (t_1 - x_3)^{a_7} (t_2 - x_1)^{a_8} (t_2 - x_2)^{a_9} (t_2 - x_3)^{a_{10}} (t_1 - t_2)^{a_{11}} ]
  \label{DF2}
\end{equation}
where the parameters can be found from the explicit contractions between the free fields in vertex operators. To understand the dimension of the solution space, we have to check the irreducible decomposition of the representations (in this case, all are $\frac{1}{2}$).
\begin{equation}
    \frac{1}{2} \otimes \frac{1}{2}  \otimes \frac{1}{2}  \otimes \frac{1}{2}  \otimes \frac{1}{2}  \otimes \frac{1}{2} = \boxed{5 \times 0} \oplus 9 \times 1 \oplus 5 \times 2 \oplus 1 \times 3
\end{equation}
It has 5 singlets. It suggests that the dimension of the solution space is 5. Hence, the following DF integral (\ref{DF2}) could be a linear combination of 5 conformal blocks.
\begin{equation}
    F = F_1 + F_2 + F_3 + F_4 + F_5
\end{equation}
Now, by placing such $F$ in the initial KZ equation (\ref{kz}) equation, we expect to have a 5th order differential equation with more coefficients (like those A,B,C), which will depend on the parameters ($a_{1} , \dots a_{11} $) of our integral ($a_1 , \dots a_{11}$). And in this case, these blocks ($F_1, F_2 , \dots F_5$) will no longer be a generic ${}_2F_{1}$ hypergeometric function. And a careful observation is required to derive the corresponding generalized hypergeometric functions\\
\newline
Now if we consider the $n$-point function and at every point take the vertex operator corresponding to the fundamental representation, we will have the points $ (0,1,\infty)$  and $ (x_1, \dots, x_{n-3})$. To determine the dimension of the solution space, we again need to decompose the tensor product of all these representations.
\begin{equation}
    \bigotimes_{i=1}^{n} \Lambda_i; \ \ \ \text{where } \Lambda = \frac{1}{2}
\end{equation}
And we need to determine how many singlets appear in this setup. In this simple example, this number is actually given by the Catalan numbers, which can be represented as matrix integrals over $SU(2)$. Here, it is important to note that singlets are produced only when our $n$-point function is even, that is, when $n = 2l$, then
\begin{equation}
    \text {dim \{solution space\} }= \text{dim}(F) = \frac{(2l)!}{(l+1) (l!)^2}
\end{equation}
But when we consider the arbitrarily representation of $sl(2)$ or even $sl(N)$ in each vertex operator, the multiplicity of the trivial representation in the irreducible decomposition (number of singlets) can be investigated by character expansionы\cite{KoikeTerada, Koike}. \\
\newline
So in this case, our DF integral will be the sum $ dim(F)$ independent conformal blocks
\begin{equation}
    F = \sum_{i=1}^{\text{dim}(F)} F_i
    \label{cb1}
\end{equation}
Now, this particular linear combination of blocks (\ref{cb1}) solves the KZ equation (\ref{kz}). The conformal block takes the form of the ${}_2F_{1}$ hypergeometric function only in the case of the four-point function. As we increase the number of insertions in the correlator, the blocks are no longer given by such an ordinary hypergeometric function. It will take some generalized hypergeometric integral form. \\
\newline
Now, let us return to the question of the number of contours, the general form of the DF integral, and the expected number of parameters in the integrand. In this specific example, when the vertex operators at all points are labeled by the same fundamental representation of $sl(2)$,
\begin{equation}
    \text{number of close contours} \oint dt_1, \dots\oint dt_r \   \text{will be:} \ \ r =  \frac{n-2}{2}
\end{equation}
Within this framework, we have three degrees of freedom: the choice of an n-point function, the use of an arbitrary representation (beyond the fundamental), and the generalization of the underlying symmetry algebra from $sl(2)$ to $sl(N)$ for the corresponding affine algebra. This comprehensive viewpoint enables us to consider such $n$-point functions for vertex operators labeled with arbitrary representations of higher algebra. \\
\newline
In that case, the general form of the DF integral take the following form:
\begin{equation}
\begin{aligned}
  & F =   \Bigl \langle \tilde{V}_{\Lambda_1, \mu_1} (0)  V_{\Lambda_2, \mu_2} (1) V_{\Lambda_3, \mu_3}(\infty) V_{\Lambda_4, \mu_4} (x_1) \dots V_{\Lambda_n, \mu_n} (x_{n-3})  V_s(\infty) Q_1(t_1) \dots Q_{r}(t_r) \Bigr \rangle  = \\ & =   \oint dt_1 \cdots \oint dt_{r} \prod_{i=1}^r \left[ t_i^{P_i} (1 - t_i)^{Q_i} \prod_{j=1}^{n-3} (t_i - x_j)^{R_{i,j}} \right] \prod_{1 \leq i < k \leq r} (t_i - t_k)^{D_{i,k}}
    \end{aligned}
\end{equation}
with parameters $P_i, Q_i, R_{i,j}$ and $D_{i,k}$. This integral representation of the $n-$point function looks similar to the original Dotsenko-Fateev integrals\cite{DF}, but not exactly the same in the powers in the integrand. \\
\newline
In the following diagram in Figure \ref{fig22}, we present this general relation of algebra, representations and multipoint correlator in a graphical presentation.\\

\newpage
Here we consider the following triplet (algebra, representation, multipoint correlator). At the center of this diagram, sits the  simplest case: the 4-point correlator of $\hat{sl}(2)_k$ WZW model when all the 4 vertex operator are labeled by the fundamental representation. Then, when we move in different directions from the center, there might be several possible cases.
\begin{itemize}
    \item  [1.] 4-point correlator of $\hat{sl}(2)_k$ with any representation at each point or different representations at different points.
    \item [2.] n-point correlator of $\hat{sl}(2)_k$ with the fundamental representation at each point.
    \item [3.] n-point correlator of $\hat{sl}(2)_k$ with any representation at each point or different representations at different points.
    \item [4.] n-point correlator of $\hat{sl}(N)_k$ with the fundamental representation at each point.
    \item  [5.] n-point correlator of $\hat{sl}(N)_k$ with any representation at each point or different representations at different points.
\end{itemize}

Throughout this paper, we have focused only on the central point and the left middle point of this diagram. A careful treatment of all these points deserves further attention, in particular the explicit computation of the conformal blocks and the demonstration of their equivalence to KZ solutions. The free-field approach for the WZW model provides a practical framework for such an investigation. Along these lines, an explicit derivation of correlators involving not only primary fields but also their descendants may be especially promising for future developments.
\section{Conclusion}

In this paper, we made the first steps towards developing an efficient calculus of conformal blocks in WZW model with the help of the free-field representation from \cite{GMMOS}. In this approach, the answers are correlators of free fields, coming from vertex operators and screening charges, and have the form of hypergeometric-type integrals (made from elliptic and higher theta functions on surfaces of non-trivial genus). But screenings convert these simple expressions into far less trivial multiple integrals. At the same time, these integrals should satisfy Knizhnik-Zamolodchikov (KZ) equations, which is a somewhat non-trivial check. In fact, there is a whole zoo of conformal blocks and they depend on:

\begin{itemize}
    \item[--] the affine (Kac-Moody) algebra $\hat G$
    \item[--] representations
    \item[--] the level of descendants
    \item[--] the number of punctures and the genus of the Riemann surface
\end{itemize}

Moreover, there are selection rules, inherited from the classical symmetry $G$. In this way, the true WZW correlators, which are certain bilinear combinations of conformal blocks, are constructed. Thus, there are plenty of things to calculate and consistency checks to be performed.\\

The whole story is long and requires many steps in many directions. This paper mostly sets the profile, reminds the basics of the free field calculus, largely forgotten during the last years, introduces the notation and presents some examples. We explained the ways to check global symmetries and KZ equations, and did it in the examples -- for $ \hat{sl}(2)_k$ and $\hat{sl}(3)_k$ WZW model and with at most double integrals. Still, this is already more than just single hypergeometric integrals, and it helps to demonstrate the essence of the generic problem.

\section*{Acknowledgements}

We are indebted to Pavel Suprun, Maxim Reva, Andrey Mironov, Elena Lanina, Konstantin Pushkin, and Nikita Kolganov for detailed discussions and comments. HS acknowledges Alexey Litvinov for helpful discussions on this study. We thank the referee of this paper for numerous important remarks and corrections. The work was partially funded within the state assignment of the Institute for Information Transmission Problems of RAS. HS was supported by the scholarship from the Theoretical Physics and Mathematics Advancement Foundation «BASIS». \\

\end{document}